\documentclass[%
 reprint,
nofootinbib,
nobibnotes,
 amsmath,amssymb,
 aps, prd,
 longbibliography,
floatfix,
]{revtex4-2}

\usepackage{graphicx}
\usepackage{dcolumn}
\usepackage{bm}
\usepackage[normalem]{ulem}

\usepackage[english]{babel}
\usepackage{fancyhdr}
\usepackage{amsmath}
\usepackage{amssymb}
\usepackage{amsfonts}
\usepackage{psfrag}
\usepackage[applemac]{inputenc}
\usepackage[bf,footnotesize]{caption}
\usepackage[dvipsnames]{xcolor}
\usepackage[colorlinks=True, citecolor=blue, linkcolor=blue, urlcolor=blue,linktocpage]{hyperref}
\usepackage{amsthm}
\usepackage{gensymb}
\usepackage{subfig}
\usepackage{physics}
\usepackage{array}
\usepackage{tcolorbox, mathtools}
\usepackage{soul}
\tcbuselibrary{skins}
\newtcolorbox{mybox}[1][]{before=\centering, drop fuzzy shadow, enhanced, colframe=blue, fonttitle=\bfseries, title=#1, center title}
\usepackage{aas_macros}

\newcommand{\mpl}{M_\mathrm{P}}

\begin{document}

\title{Angular momentum sensitivities in scalar-tensor theories}

\author{Adrien Kuntz}
 \email{adrien.kuntz@sissa.it}
\author{Enrico Barausse}
\email{barausse@sissa.it}

\affiliation{SISSA, Via Bonomea 265, 34136 Trieste, Italy and INFN Sezione di Trieste}
\affiliation{IFPU - Institute for Fundamental Physics of the Universe, Via Beirut 2, 34014 Trieste, Italy}

\begin{abstract}
    Scalar-tensor theories have a long history as possible phenomenological alternatives to General Relativity, but are known to potentially produce deviations from the (strong) equivalence principle in systems involving self-gravitating objects, as a result of the presence of an additional gravitational scalar field besides the tensor modes of General Relativity.
We describe here a novel mechanism whereby the  equivalence principle is violated  
for an isolated rotating neutron star, if the  gravitational scalar field is changing in time far from the system. We show that the neutron star rotational period changes due to an effective coupling (``angular momentum sensitivity'') to the  gravitational scalar, and compute that coupling for viable equations of state for nuclear matter. We comment on the relevance of our findings for testing scalar-tensor theories and models of ultralight dark matter with pulsar timing observations, a topic that we tackle in a companion paper.
\end{abstract}

\date{\today}

\maketitle

\section{Introduction}
Scalar-tensor theories of gravity have a long history as possible phenomenological alternatives to General Relativity (GR), dating back to seminal work by Fierz~\cite{Fierz:1939ix}, Jordan~\cite{Jordan:1959eg}, Brans and Dicke~\cite{Brans:1961sx,Dicke62} (FJBD). This class of theories, hereafter referred to as FJBD-like theories, describe gravitation by a scalar degree of freedom alongside the usual tensor polarizations of GR. Historically, the theory has been defined in the  Jordan frame, where the gravitational scalar is minimally coupled to matter and conformally coupled to the usual graviton tensor field.  This frame choice  makes it apparent that at tree level
the scalar field does not produce any fifth forces (because of the absence of direct couplings to matter). However, the scalar can produce deviations from the equivalence principle in strongly gravitating systems (e.g. neutron stars), through non-universal scalar interactions~\cite{Nordtvedt68,Eardley1975ApJ,1989ApJ...346..366W,1977ApJ...214..826W,Damour:1992we,Will:1993ns}.

This violation of the ``strong'' equivalence principle is ripe of consequences for the dynamics of binaries of compact objects in scalar-tensor theories. In more detail, 
the effective scalar force between two objects can be parametrized by scalar charges (or ``sensitivities''), which describe a body's response to a change in the local scalar field amplitude (which in turn renormalizes the local value of the Newton constant and thus the body's gravitational binding energy and mass)~\cite{Nordtvedt_1968,Nordtvedt:1968qs,Eardley1975ApJ,1989ApJ...346..366W,1977ApJ...214..826W,Damour:1992we}. It has long been recognized that these sensitivities cause the appearance of monopole and/or dipole gravitational wave (GW) fluxes from binaries of compact objects~\cite{Eardley1975ApJ,1977ApJ...214..826W,Damour:1992we}, which dominate the dynamics and evolution of these systems at low frequencies. Indeed, these scalar fluxes are also present in scalar-tensor theories more general than the 
FJBD ones, belonging to the Horndeski~\cite{Horndeski:1974wa}, beyond Horndeski~\cite{Gleyzes:2014dya} and DHOST classes~\cite{BenAchour:2016fzp},
in which the generation of GWs from compact binaries is starting to be investigated.

The presence of monopole and dipole fluxes is crucial to establish experimental viability of scalar-tensor theories, since no evidence of the latter is present in  binary pulsar systems~\cite{Damour:1991rd,weisberg2004relativistic,Kramer:2006nb} and in the binaries of 
neutron stars and black holes observed by ground interferometers~\cite{LIGOScientific:2018dkp,LIGOScientific:2019fpa,LIGOScientific:2020tif}. Indeed, neutron star/pulsar binaries have severely constrained the parameter space of
FJBD-like scalar-tensor theories~\cite{Damour_1998,Freire:2012mg,Shao:2017gwu}. The same may happen 
for more general Horndeski, beyond Horndeski and DHOST theories, provided that the generation of GWs in these theories is theoretically understood, despite the presence of non-linear screening mechanisms~\cite{deRham:2012fw,Dar:2018dra,Kuntz:2019plo,Bezares:2021dma,Boskovic:2023dqk}.

Besides their interest as phenomenological alternatives to GR, as well as possible effective field theories of Dark Energy and/or inflation~\cite{Cheung:2007st, Gubitosi:2012hu}, scalar-tensor theories can also be used to describe, in some regimes, the physics of axion-like particles. The latter are enjoying growing popularity as a model of ``fuzzy'' Dark Matter~\cite{Hui:2016ltb}, which could address several shortcomings of the cold Dark Matter paradigm on small scales, provided that the mass of the particle is sufficiently small ($m_{\varphi}\sim 10^{-22}$ eV). These models are therefore also known as ultralight Dark Matter, and on the scales of galaxies and compact objects they can be effectively represented as scalar-tensor theories (with the presence of a potential accounting for the scalar mass $m_{\varphi}$). Interestingly, although crucial for the cosmological viability of the theory, the scalar mass $m_\varphi$ can be neglected on the local scales of compact objects, which are generally way smaller than the axion-like particle's Compton wavelength.

Driven by these reasons, in this paper we will consider the class of scalar-tensor theories of the
FJBD type, 
with a  conformal coupling chosen to be linear -- as in the original FJBD theory -- or more general, e.g. quadratic -- as in Damour-Esposito-Far\`{e}se (DEF) gravity~\cite{Damour:1992we,PhysRevLett.70.2220}. Both possibilities have  been investigated in the literature on fuzzy Dark Matter (see e.g.~\cite{Blas:2016ddr,Blas:2019hxz}). Moreover, we will assume that the scalar field is  varying in time on galactic/cosmological scales as a result of a small mass $m_\varphi$, which can be neglected in the local dynamics, but which can produce scalar field oscillations with frequency $m_\varphi$ on large scales. These oscillations have indeed been used to constrain models of fuzzy Dark Matter with pulsar-timing arrays, since the scalar field changes
would produce time variations in the Galactic Newtonian potential, which would in turn cause the appearance of an integrated Sachs-Wolfe timing residual in pulsar data~\cite{Khmelnitsky:2013lxt,Porayko:2018sfa,EuropeanPulsarTimingArray:2023egv,NANOGrav:2023hvm}. 

Here, we propose a novel way to use pulsar data to test the presence of a possible ultralight gravitational scalar field. 
In more details, the basic idea, which we will develop below, is that the conformal coupling renormalizes the local value of the Newton constant
experienced by a pulsar. This in turn changes the pulsar gravitational binding energy, and therefore the pulsar's moment of inertia
and  rotational frequency. Indeed, this change is modulated by the scalar field oscillations far from the pulsar, 
which lie in the nHz band of pulsar-timing arrays for 
$m_{\varphi}\sim 10^{-22}$ eV. In this paper, we will develop the calculation of this effect from first principles, and show that it can be parametrized by angular momentum ``sensitivies'' (or ``charges'') akin to those describing the scalar fluxes from compact binaries. In a companion paper~\cite{Smarra:2024kvv}, we will apply the idea to data from the European Pulsar Timing Array~\cite{2013CQGra..30v4009K} to constrain the theory.

The paper is organized as follows. In Section~\ref{sec:framework} we will introduce the scalar-tensor theories which we study in this article and derive the equations of motion for a slowly rotating neutron star model. Section~\ref{sec:asymptot_obs_sensitivities} contains the major new theoretical results of this article: defining asymptotically measured parameters of the rotating neutron star (Section~\ref{sec:jordan}) and relating them to conserved quantities (Section~\ref{sec:conserved}) which turn out to be different in scalar-tensor theories. This allows us to define in Section~\ref{sec:sensitivities} the angular momentum sensitivity in a similar way as the well-known mass sensitivity. Interestingly, the Newtonian limit of the angular momentum sensitivity is nontrivial and depends on the equation of state (EOS) considered, as we discuss in Section~\ref{sec:newtonian}.
Section~\ref{sec:results} finally contains our numerical results for the mass and angular momentum sensitivities, both for FJBD and DEF theories.


\section{Framework} \label{sec:framework}

\subsection{Theory} \label{sec:theory}

We consider a  scalar-tensor theory with Einstein frame action given by
\begin{equation}
    S = \mpl^2 \int \mathrm{d}^4x \sqrt{-g} \, \bigg( \frac{R}{2} - g^{\mu \nu} \partial_\mu \varphi \partial_\nu \varphi \bigg) + S_m[ \psi_m, \tilde g_{\mu \nu}] \; ,
\end{equation}
where $S_m$ is the matter action, the scalar field is dimensionless and the Planck mass is $\mpl^2 = 1/8\pi G_b$. We denote the bare Newton constant by $G_b$  to avoid confusion with the value of the gravitational constant  \textit{measured}  by local experiments, which will be defined in the following.
The Jordan-frame metric 
in which matter 
does not couple directly to the scalar (i.e. the metric in
which matter follows geodesics)
is given by $\tilde g_{\mu \nu} = A^2(\varphi) g_{\mu \nu}$. We will denote all quantities in the Jordan frame with a tilde. The action can also be equivalently expressed in the Jordan frame as
\begin{align}
    S &= \mpl^2 \int \mathrm{d}^4x \sqrt{- \tilde g} A^{-2}(\varphi) \times \nonumber \\ \times\bigg[ \frac{\tilde R}{2} 
    &- \big(1-3\alpha^2(\varphi)\big) \tilde g^{\mu \nu} \partial_\mu \varphi \partial_\nu \varphi \bigg] + S_m[ \psi_m, \tilde g_{\mu \nu}] \; ,
\end{align}
where
\begin{equation} \label{eq:logderA}
    \alpha(\varphi) = A'(\varphi) / A(\varphi) \; .
\end{equation}

The conformal factor will be chosen to be either $A(\varphi) = e^{\alpha \varphi}$ (FJBD theory~~\cite{Fierz:1939ix,Jordan:1959eg,Brans:1961sx,Dicke62}) or $A(\varphi) = e^{ \beta \varphi^2/2}$ (DEF spontaneous scalarization theory~\cite{PhysRevLett.70.2220}).\footnote{In the first case, we are committing a slight abuse of notation, because we denote the fundamental coupling constant of the theory, $\alpha$, with the same name as the log-derivative given by Eq.~\eqref{eq:logderA}. This is motivated by the fact that the two are equal for this theory.}
This scalar coupling $\alpha$ is also related to the historical FJBD parameter $\omega_\mathrm{FJBD}$ via $\alpha = (3+2 \omega_\mathrm{FJBD})^{-1/2}$~\cite{Barausse:2012da,Palenzuela:2013hsa}.  
All the results derived in this article also hold for an ultralight scalar (e.g. an axion-like Dark Matter field), provided that the Compton wavelength $\lambda = 1/m_\varphi$ of the field is much larger than the size of a neutron star.

From an experimental point of view, these scalar-tensor theories are both already quite constrained by observations. Indeed, in the first case, the best experimental constraint on the $\alpha$ parameter comes from the Cassini mission~\cite{Bertotti:2003rm}, which requires $\alpha^2 \lesssim 10^{-5}$. In the second case, the absence of
significant 
deviations from the GR predictions in binary pulsar data requires $\beta \gtrsim -4.3$ 
(depending on the exact EOS), while the value $\varphi_0$ of the scalar field
on cosmological scales is restricted to be $|\beta \varphi_0| \lesssim 3 \times 10^{-4}$~\cite{Freire:2012mg, Shao:2017gwu}. 

In FJBD-like scalar-tensor theories the value of the Newton constant measured by local observers, $\tilde G$, depends on the 
local value  $\varphi_0$ of the field ``felt'' by observer
through~\cite{1989ApJ...346..366W,Damour:1992we}
\begin{equation} \label{eq:Gtilde}
    \tilde G = G_b A^2(\varphi_0) \big( 1+\alpha^2(\varphi_0) \big) \; .
\end{equation}
For a neutron star, $\varphi_0$ corresponds to the 
 the cosmological value of the scalar far from the system.
This  dependence of $\tilde G$ on the scalar field led Nordtvedt~\cite{Nordtvedt_1968,Nordtvedt:1968qs}, Eardley~\cite{Eardley1975ApJ} and Will~\cite{1977ApJ...214..826W} to suggest that scalar-tensor theories violate the strong equivalence principle, because the inertial mass of a massive body depends on the local value of $\tilde G$ through its dependence on the gravitational energy. 

The inertial mass $M(\varphi_0)$ is defined by ``skeletonizing'' the action for a massive object, i.e. by replacing the matter action by a simpler point-particle effective action:
\begin{equation} \label{eq:inertialMass}
    S_m[\psi_m, \tilde g_{\mu \nu}] \rightarrow S_\mathrm{PP} = - \int \mathrm{d}\tilde \tau \, M(\varphi_0) \; ,
\end{equation}
where $\mathrm{d}\tilde \tau =  \sqrt{-\tilde g_{\mu \nu} \mathrm{d}x^\mu \mathrm{d}x^\nu}$ is the proper time of the point-particle in the Jordan frame. 
This dependence of the mass on the scalar field is encoded in the  (mass) ``sensitivities'' $s_M$, which represent the response of the inertial mass to changes in the scalar field, i.e.~\footnote{Notice that this derivative is slightly different from the derivative with respect to $\ln \tilde G$, because of the $1+\alpha^2$ term in Eq.~\eqref{eq:Gtilde}.}
\begin{equation} \label{eq:sensitivity}
    s_M = - \frac{1}{2} \frac{\mathrm{d} \ln M}{\mathrm{d} \ln A} \bigg |_N =  - \frac{1}{2 \alpha} \frac{\mathrm{d} \ln M}{\mathrm{d} \varphi_0} \bigg |_N \; ,
\end{equation}
where the baryon number of the star, $N$, is kept constant. 
For weakly self-gravitating bodies, $s_M \rightarrow 0$, since in that limit the contribution of the gravitational binding energy to the mass (and thus the dependence on $\tilde G$) disappear. However, for a strong-field object such as a neutron star, the sensitivities can become of order one (or even larger, when spontaneous scalarization happens~\cite{PhysRevLett.70.2220}, c.f. section~\ref{sec:results}).

If the mass depends on the asymptotic value of the scalar, so does the moment of inertia $I$ of the neutron star. We will thus define an angular momentum sensitivity for a neutron star as
\begin{equation} \label{eq:sIintro}
    s_I = - \frac{1}{2 \alpha}  \frac{\mathrm{d} \ln I}{\mathrm{d} \varphi_0} \bigg |_N \; .
\end{equation}
We will see in section~\ref{sec:sensitivities} that  knowledge of this quantity can be used to determine the variation in the observed frequency of a pulsar caused by changes in  the asymptotic scalar value. In particular, our results will be relevant for testing scalar-tensor theories and models of ultralight Dark Matter with pulsar timing observations, a topic that we tackle in a companion paper.

Although moments of inertia for rotating neutron stars have recently been computed in several previous works~\cite{Pani:2014jra, Yazadjiev:2016pcb, Silva:2014fca, Doneva:2018ouu, Doneva:2016xmf, Popchev:2018fwu}, we are only aware of a few older articles where the angular momentum sensitivities are mentioned~\cite{1989ApJ...346..366W, Will:1993ns, 1990MNRAS.244..184G,Damour:1996ke}. In this work, we will thus update the computed values of angular momentum sensitivities for both theories using modern equations of state for nuclear matter. We will also clarify the weak field limit of the angular momentum sensitivities, and relate the latter to the EOS in section~\ref{sec:newtonian}.

\subsection{Neutron star model and equations of motion in Einstein frame}

The energy-momentum tensor of the    neutrons star matter is modeled as a perfect fluid, defined in the Jordan frame as
\begin{equation}
    \tilde T^{\mu \nu} = (\tilde \epsilon + \tilde p) \tilde u^\mu \tilde u^\nu + \tilde p \tilde g^{\mu \nu} \; ,
\end{equation}
where $\tilde \epsilon$ and $\tilde p$ are the energy density and pressure of the fluid (naturally defined in the Jordan frame), while $\tilde u^\mu$ is the 4-velocity of the fluid, normalized to satisfy $\tilde g_{\mu \nu} \tilde u^\mu \tilde u^\nu = -1$. In the Einstein frame, one can find the energy-momentum tensor of matter as $T^{\mu \nu} = A^6(\varphi) \tilde T^{\mu \nu}$ from the relations $T^{\mu \nu} = 2/\sqrt{-g} \, \delta S_m/\delta g_{\mu \nu}$, $\tilde T^{\mu \nu} = 2/\sqrt{-\tilde g} \delta S_m/\delta \tilde g_{\mu \nu}$. Similarly, $u^\mu = A(\varphi) \tilde u^\mu$.
We assume an axisymmetric and stationary metric ansatz. As commented in the introduction, this assumption is valid provided that the asymptotic value of the scalar, $\varphi_0$, varies on timescales much larger than the pulsar period, so that we can consider $\varphi_0$ as  constant in our setup. We then adiabatically vary $\varphi_0$ from one equilibrium solution to the next in order to obtain results for a slowly evolving scalar field.

We assume a uniformly rotating fluid configuration with frequency $\Omega = \tilde u^\phi / \tilde u^t = u^\phi/u^t$, which is a good approximation for pulsars~\cite{2013rehy.book.....R}.
This last equation suggests that $\Omega$ is frame-invariant, so we will omit the tilde when referring to this quantity.
However, at this point a clarification is needed in order to interpret $\Omega$ as the rotational frequency of the fluid. Indeed, notice that this definition of $\Omega$ as the ratio of two vector components depends on the specific time and angular coordinates chosen to describe the fluid. Thus, while $\Omega$ is \textit{frame-invariant}, it is not \textit{coordinate-invariant}. This distinction is important, because we will see  in section~\ref{sec:jordan} that a change of coordinates is necessary in order to define asymptotically measured quantities. Conversely, notice that $\tilde p$ and $\tilde \epsilon$ are scalars (coordinate-invariant), but they transform as $p = A^4(\varphi) \tilde p$, $\epsilon = A^4(\varphi) \tilde \epsilon$ when changing frame.


Computations greatly simplify if we focus on the slowly rotating limit and consider only perturbations linear in $\Omega$, on top of a nonrotating configuration. In this case, one can write the metric in spherical coordinates as~\cite{1967ApJ...150.1005H}
\begin{align} \label{eq:einstframemetric}
    &g_{\mu \nu} \mathrm{d}x^\mu \mathrm{d}x^\nu = - e^{\nu(r)} \mathrm{d}t^2 + \frac{\mathrm{d}r^2}{1-\frac{2\mu(r)}{r}} \nonumber \\
    &+ r^2 \big( \mathrm{d}\theta^2 + \sin^2\theta \mathrm{d}\phi^2 \big) 
    + 2(\omega(r) - \Omega) r^2 \sin^2 \theta \mathrm{d} \phi \mathrm{d}t \; ,
\end{align}
and the velocity of the fluid in the Einstein frame is $u^\mu = e^{-\nu/2} (1, 0, 0, \Omega)$. 
From the Einstein equations and the stress-energy tensor conservation $\tilde \nabla_\mu \tilde T^{\mu \nu} =0$, one obtains the following system of equations~\cite{PhysRevLett.70.2220, Pani:2014jra, Yazadjiev:2016pcb}:
\begin{align} \label{eq:EOMmu}
    \mu ' &= 4 \pi G_b r^2 A^4(\varphi) \tilde \epsilon + r(r-2\mu) \frac{\psi^2}{2} \; , \\
    \nu' &= 8 \pi G_b \frac{r^2 A^4(\varphi)}{r-2\mu} \tilde p + r \psi^2 + \frac{2 \mu}{r(r-2\mu)} \; ,\\
    \varphi ' &= \psi \; ,\\
    \psi' &= 4 \pi G_b \frac{r A^4(\varphi)}{r-2\mu} \bigg[ \alpha(\varphi) (\tilde \epsilon-3\tilde p) + r \psi (\tilde \epsilon - \tilde p) \bigg] \nonumber \\
    &- 2 \frac{r-\mu}{r(r-2\mu)} \psi \; , \\
    \tilde p' &= - (\tilde \epsilon+\tilde p) \bigg[  4 \pi G_b \frac{r^2 A^4(\varphi)}{r-2\mu} \tilde p + r \frac{\psi^2}{2} \nonumber \\
    &+ \frac{\mu}{r(r-2\mu)} + \alpha(\varphi) \psi \bigg] \; , \\
    \kappa &= \omega' \; , \\
    \kappa' &= 4 \pi G_b A^4(\varphi) \frac{r}{r-2\mu} (\tilde \epsilon + \tilde p) \big( r \kappa + 4 \omega \big) \nonumber \\
    &+ r \psi^2 \kappa - \frac{4}{r} \kappa \; . \label{eq:EOMkappa}
\end{align}
Note that the equations for $\mu, \varphi, \psi$ and $\tilde p$ form a closed subsystem, while $\nu$ and $\omega$ can be obtained by  integrating their equation of motion, once the other variables are known. To close the system, we need an EOS $\tilde \epsilon = \tilde \epsilon(\tilde p)$ relating pressure and energy density for the 
nuclear matter of the neutron star. Different models for this EOS are available. In this work, we adopt three representative EOS in agreement with
the observation of GW170817~\cite{the_ligo_scientific_collaboration_gw170817:_2017}, namely \texttt{AP4}, \texttt{SLy} and \texttt{MPA1} (see e.g.~\cite{Read:2008iy} for a discussion of these EOS models). 

We then have to numerically integrate these equations. We do so in two steps. First, we integrate the system from $r=0$ to the radius of the star $r=r_S$, defined by the vanishing  pressure condition $\tilde p(r_S) = 0$. The second step consists of integrating the equations in vacuum ($\tilde \epsilon = \tilde p = 0$) from the surface $r_S$  up to large distances, where one can analyze the asymptotic behavior of the metric and  scalar field to extract physically relevant quantities.

The  boundary conditions at the center of the star are obtained
by imposing regularity, which leads to the conditions
\begin{align}
    \mu(0) &= 0 \; , \quad \nu(0) = \nu_c \; , \quad \varphi(0)= \varphi_c \; , \quad \psi(0) = 0 \; , \nonumber \\
    \quad \tilde p(0) &= p_c \; , \quad \omega(0) = \omega_c \; , \quad \kappa(0) = 0 \; .
\end{align}
 One can then note that the system of Eqs.~\eqref{eq:EOMmu}-\eqref{eq:EOMkappa} is linear in $\omega$ and invariant under constant shifts of $\nu$. We use this freedom to set the values of these quantities at the center to
\begin{equation}
     \nu(0) = 0 \; , \quad \omega(0) = 1 \; .
\end{equation}
Once the system is numerically integrated, one can again shift $\nu$ and rescale $\omega$ in order to ensure asymptotic flatness for the Einstein-frame metric, i.e. $\nu(\infty) = 0$, $\omega(\infty) = \Omega$. 
Thus, the only boundary conditions at $r=0$ that we need for integrating the system are the values of the central pressure $p_c$ and scalar $\varphi_c$. Equivalently, this means that we have a two-parameter family of solutions, which we can describe more conveniently by measured quantities such as the mass and the scalar field at infinity. In the Einstein frame, we thus define the gravitational mass $M_E$, the dimensionless scalar charge $q$ and the angular momentum $J_E$ as the asymptotic values appearing in the Einstein frame metric~\eqref{eq:einstframemetric} near spatial infinity, i.e.:
\begin{align}
    \nu &= - \frac{2 G_b  M_E}{r} + \mathcal{O} \bigg( \frac{1}{r^2} \bigg) \label{eq:einstframemass} \; , \\
    \varphi &= \varphi_0 - q \frac{G_b  M_E}{r} + \mathcal{O} \bigg( \frac{1}{r^2} \bigg) \label{eq:defq} \; ,\\
    \omega &= \Omega - \frac{2 G_b J_E}{r^3}  + \mathcal{O} \bigg( \frac{1}{r^4} \bigg) \label{eq:einstframeJ} \; .
\end{align}
From the equation of motion for $\omega$~\eqref{eq:EOMkappa}, one can also obtain a convenient integral expression for the Einstein frame angular momentum (see e.g.~\cite{Yazadjiev:2016pcb}):
\begin{equation} \label{eq:integralJ}
    J_E = \frac{8 \pi}{3} \int_0^{r_S} r^4 A^4(\varphi) \big( \tilde \epsilon + \tilde p \big) \frac{e^{-\nu/2}}{\sqrt{1-\frac{2 \mu}{r}}} \omega \,  \mathrm{d}r \; ,
\end{equation}
where we stress that no gravitational constant appears in this equation (this is obvious from dimensional analysis).

From the mass in Einstein frame $M_E$, we can easily obtain the inertial mass $M$ defined in Eq.~\eqref{eq:inertialMass}  by performing a frame transformation, yielding $M_E = A(\varphi_0) M$. This allows for computing the mass sensitivity $s_M$ from Eq.~\eqref{eq:sensitivity}, c.f. section~\ref{sec:sensitivities} for more details.
In addition, let us stress there is a well-known relation between the sensitivity and the scalar charge $q$~\cite{Damour:1992we,Palenzuela:2013hsa}:
\begin{equation} \label{eq:relqsm}
    q = \alpha(1-2s_M) \; .
\end{equation}
We prove this relation in Appendix~\ref{app:scalcharge}.

\section{Asymptotic observables and sensitivities} \label{sec:asymptot_obs_sensitivities}

\subsection{Measured quantities in the Jordan frame} \label{sec:jordan}

The asymptotic quantities mentioned in the previous subsection, which we obtain after numerical integration, are defined in the Einstein frame. However, because matter follows geodesics of the metric in the Jordan frame, the latter is the frame in which observed quantities must be defined. For instance, if a distant observer measures the mass of an object using Kepler's law (i.e. geodesics), that mass is encoded in the asymptotic
structure of the Jordan-frame metric.
Therefore, one needs to transform to the Jordan frame in order to extract  physical quantities such as, for instance, the measured gravitational mass and angular momentum. The frame transformation (i.e. field redefinition) reads
\begin{align}
       &\tilde g_{\mu \nu} \mathrm{d}x^\mu \mathrm{d}x^\nu = A^2(\varphi) \bigg[ - e^{\nu} \mathrm{d}t^2 + \frac{\mathrm{d}r^2}{1-\frac{2\mu}{r}} \nonumber \\
    &+ r^2 \big( \mathrm{d}\theta^2 + \sin^2\theta \mathrm{d}\phi^2 \big) 
    + 2(\omega - \Omega) r^2 \sin^2 \theta \mathrm{d} \phi \mathrm{d}t \bigg] \;.
\end{align}
From this equation it follows that, in order to ensure asymptotic flatness, one has to rescale the time and radial coordinates. This coordinate transformation is
\begin{equation}
    \tilde r = A(\varphi_0) r \; , \quad \tilde t = A(\varphi_0) t \; ,
\end{equation}
where we recall that $\varphi_0$ is the asymptotic value of the scalar field. In the new coordinates the metric reads
\begin{align} \label{eq:jordanframemetric}
       &\tilde g_{\mu \nu} \mathrm{d}\tilde x^\mu \mathrm{d}\tilde x^\nu = \frac{A^2(\varphi)}{A^2(\varphi_0)} \bigg[ - e^{\nu} \mathrm{d}\tilde t^2 + \frac{\mathrm{d}\tilde r^2}{1-\frac{2\mu A(\varphi_0)}{\tilde r}} \nonumber \\
    &+ \tilde r^2 \big( \mathrm{d}\theta^2 + \sin^2\theta \mathrm{d}\phi^2 \big) 
    + 2\frac{\omega - \Omega}{A(\varphi_0)} r^2 \sin^2 \theta \mathrm{d} \phi \mathrm{d}t \bigg] \; .
\end{align}
Additionally, frequencies get rescaled by the coordinate transformation as one moves from the Einstein frame to the Jordan frame. Thus, with our previous choice of coordinates, $\Omega$ is \textit{not} the frequency of the fluid as measured by distant observers: the correct measured frequency is $\Omega_\mathrm{obs}= \Omega / A(\varphi_0)$. 

Using the Jordan-frame metric in the new coordinates, we define the (Jordan-frame) gravitational mass and angular momentum of the system from the asymptotic limit
\begin{align}
    e^{\nu} \frac{A^2(\varphi)}{A^2(\varphi_0)} &= 1 - \frac{2 \tilde G \tilde M}{\tilde r} + \mathcal{O} \bigg( \frac{1}{\tilde r^2} \bigg) \; , \label{eq:tildeMmetric} \\ 
    \frac{\omega - \Omega}{A(\varphi_0)} &= - \frac{2 \tilde G \tilde J}{\tilde r^3}  + \mathcal{O} \bigg( \frac{1}{\tilde r^4} \bigg) \; ,
\end{align}
where we recall that the measured value of the Newton constant $\tilde G$ is given in Eq.~\eqref{eq:Gtilde}. It is then straightforward to derive the relation between Jordan and Einstein frame asymptotic quantities by inserting Eqs.~\eqref{eq:einstframemass}-\eqref{eq:einstframeJ} into the Jordan frame metric~\eqref{eq:jordanframemetric}:
\begin{align}
    \tilde J &= \frac{J_E}{1+\alpha^2} \label{eq:tildeJ} \; , \\
    \tilde M &= \frac{M_E}{A(\varphi_0)} \frac{1+\alpha q}{1+\alpha^2} = M \frac{1+\alpha q}{1+\alpha^2} \; ,
\end{align}
where we recall that $M$ is the \textit{inertial} mass of the star, from which the mass sensitivity is defined in Eq.~\eqref{eq:sensitivity}.
From the last equation, we see that the gravitational mass differs in general from the inertial mass, which is a manifestation of the violation of the equivalence principle in scalar-tensor theories (see e.g.~\cite{Damour:1992we,1989ApJ...346..366W} for similar equations and discussion). However, for a weakly self-gravitating object, $s_M \rightarrow 0$, and we see from Eq.~\eqref{eq:relqsm} that $q \simeq \alpha$, so that $\tilde M \simeq M$. Thus, only the strong equivalence principle is violated in these theories.

From Eq.~\eqref{eq:tildeJ}, we see that the Jordan frame angular momentum also differs from its Einstein frame counterpart. In the next  section, we will show that only the \textit{Einstein frame} angular momentum is conserved when the scalar fluctuates.

\subsection{Conserved quantities} \label{sec:conserved}

In GR, the gravitational mass is a conserved quantity. However, this is not necessarily the case in scalar-tensor theories. As discussed in Sec.~\ref{sec:theory}, if the boundary value $\varphi_0$ of the scalar changes adiabatically, then the value of $\tilde G$ will also change. This changes the gravitational mass, since the latter includes the gravitational energy of the star. However, in the Jordan frame the number of baryons of the neutron star should be conserved, because non-gravitational physics is unchanged. This translates in a conservation equation $\tilde \nabla_\mu \tilde J^\mu = 0$ where $\tilde J^\mu = \tilde n_b \tilde u^\mu$ is the baryonic current density, $\tilde n_b$ being the proper baryon density of the neutron star (related to pressure and energy density by the EOS).
Thus, one can define the baryonic number of the star as
\begin{equation}
    N = \int \mathrm{d}^3 x \, \sqrt{-\tilde g} \tilde J^t = 4 \pi \int_0^{r_S} \mathrm{d}r \, r^2 \frac{\tilde n_b A^3(\varphi)}{\sqrt{1-\frac{2\mu}{r}}} \; .
\end{equation}
Note that the quantity $\mathrm{d}^3 x \, \sqrt{-\tilde g} \tilde J^t$ is obviously coordinate-invariant, but transforms under a change of frame. Therefore, we can (equivalently) define $N$ with Jordan frame coordinates $\tilde x^\mu$ (which
have a more intuitive physical meaning), or with  Einstein frame coordinates $x^\mu$ (which are more convenient for the computations).
The baryonic number will satisfy $\mathrm{d} N/ \mathrm{d}t = 0$, since
\begin{align}
\begin{split}
    \frac{\mathrm{d} N}{\mathrm{d}t} &= \int \mathrm{d}^3x \, \partial_0 \big(\sqrt{-\tilde g} \tilde J^t \big) \\
    &= - \int \mathrm{d}^3x \, \partial_i \big( \sqrt{-\tilde g} \tilde J^i \big) \\
    &= - \int \mathrm{d}^2 S_i \, \sqrt{-\tilde g} \tilde J^i \\
    &= 0 \; ,
\end{split}
\end{align}
where we have used the current conservation to go from the first line to the second, Gauss theorem to go from the second to the third, and in the last line we have used $\tilde J^r = 0$. A very important point to note is that we did not use in this proof the existence of a timelike Killing vector field: thus, we expect the baryonic mass to be conserved also in time-dependent processes involving adiabatic variations of the scalar field. This 
is in contrast with the gravitational mass, which is  defined assuming the existence of an asymptotic timelike Killing vector field.

On the other hand, conservation of angular momentum only requires the existence of
a rotational Killing vector field in the spacetime, $\xi = \partial/\partial \phi$.  
Since the adiabatic variation of 
$\varphi_0$ does not break this symmetry,
we thus expect the Komar angular momentum $J_K$ to still be conserved at fixed baryonic number,  $ \frac{\mathrm{d} J_K}{\mathrm{d} \varphi_0} \bigg |_{N} = 0$. Let us then compute the value of this conserved quantity. From the energy-momentum tensor in the Jordan frame, $\tilde T^\mu_\nu$, one can build a conserved current $\tilde J^\mu = \tilde T^\mu_\nu \xi^\nu$~\footnote{Note that the Komar definition of the current also include a term proportional to the trace of the energy-momentum tensor, which however does not contribute to the time component of the current used to define angular momentum.}. The conservation of this current follows from energy-momentum conservation and from the Killing equation in the Jordan frame. One  can then build a conserved charge in a very similar way to the baryon number:
\begin{align} \label{eq:JK}
    J_K &= \int \mathrm{d}^3 x \, \sqrt{-\tilde g} \tilde J^t \nonumber \\
    &= \frac{8 \pi}{3} \int_0^{r_S} r^4 A^4(\varphi) \big( \tilde \epsilon + \tilde p \big) \frac{e^{-\nu/2}}{\sqrt{1-\frac{2 \mu}{r}}} \omega \,  \mathrm{d}r \; ,
\end{align}
From Eq.~\eqref{eq:integralJ}, we see that the Komar angular momentum is the quantity appearing in the Einstein-frame metric, $J_K = J_E$, which differs from the Jordan-frame metric quantity $\tilde J$ by the $1+\alpha^2$ factor in Eq.~\eqref{eq:tildeJ}. 

Finally, we stress that one could also have directly defined the angular momentum in the Einstein frame by using the conservation of the {\it total} stress energy tensor, 
which includes a scalar contribution. However, the scalar does not contribute to angular momentum, because $(T_\varphi)^t_\phi=0$. This leads to the same expression for the Komar angular momentum, independently of which frame is used for the derivation.

\subsection{Sensitivities} \label{sec:sensitivities}

The mass sensitivity has been defined in Eq.~\eqref{eq:sensitivity}, and can be obtained by numerically computing the derivative at fixed baryon number. Indeed, because (non-rotating) neutron star solutions are completely determined by the scalar field $\varphi_c$ and pressure $p_c$ at the center, at fixed baryon number one has $0 = d N = \partial N / \partial \varphi_c \mathrm{d}\varphi_c + \partial N / \partial p_c \mathrm{d}p_c$, and therefore
\begin{align} \label{eq:sigmaM}
    s_M &=  - \frac{1}{2 \alpha M} \bigg( \frac{\partial N}{\partial p_c} \frac{\partial M}{\partial \varphi_c} - \frac{\partial N}{\partial \varphi_c} \frac{\partial M}{\partial p_c} \bigg) \nonumber \\
    &\times \bigg( \frac{\partial N}{\partial p_c} \frac{\partial \varphi_0}{\partial \varphi_c} - \frac{\partial N}{\partial \varphi_c} \frac{\partial \varphi_0}{\partial p_c} \bigg)^{-1} \; ,
\end{align}
where we recall that $M = M_E/A(\varphi_0)$ is the inertial mass.
We can then obtain all partial derivatives numerically, and compute the sensitivity from our numerical solutions for non-rotating stars. Our results are in good agreement with previous
works, e.g. Refs.~\cite{PhysRevLett.70.2220, Palenzuela:2013hsa}

In this work, however, we are interested in slowly rotating stars, and namely in the change of the observed frequency of a pulsar produced by an adiabatic variation of the scalar field 
at spatial infinity. To compute this effect, let us first define the moment of inertia of the pulsar as
\begin{equation} \label{eq:defI}
    I = \frac{J_K}{\Omega_\mathrm{obs}} = A(\varphi_0) \frac{J_K}{\Omega} \; .
\end{equation}
This quantity is easily computed with our numerical code by using Eq.~\eqref{eq:JK}.
By conservation of the Komar angular momentum, changes in the observed frequency are directly related to changes in the moment of inertia. We thus define an \textit{angular momentum sensitivity} $s_I$ as
\begin{equation} \label{eq:angularmomsensitivity}
    s_I = - \frac{1}{2 \alpha} \frac{\mathrm{d} \ln I}{\mathrm{d} \varphi_0} \bigg |_{N,J_K} = \frac{1}{2 \alpha} \frac{\mathrm{d} \ln \Omega_\mathrm{obs}}{\mathrm{d} \varphi_0} \bigg |_{N,J_K} \; .
\end{equation}

An important  simplification is now allowed by the fact that the equation of motion for the angular variable $\bar \omega$, Eq.~\eqref{eq:EOMkappa}, is linear. Therefore, the angular momentum scales linearly with the central value of $\bar \omega$. Because of this,  fixing the angular momentum to a constant just amounts to varying the central value of $\bar \omega$ in order to compensate for the changes in $(p_c, \varphi_c)$. As the moment of inertia is independent of the central value of $\bar \omega$ (by its very definition), we can compute $s_I$ exactly as we did for the mass sensitivity:
\begin{align} \label{eq:sigmaI}
    s_I &=  - \frac{1}{2 \alpha I} \bigg( \frac{\partial N}{\partial p_c} \frac{\partial I}{\partial \varphi_c} - \frac{\partial N}{\partial \varphi_c} \frac{\partial I}{\partial p_c} \bigg) \nonumber \\
    &\times \bigg( \frac{\partial N}{\partial p_c} \frac{\partial \varphi_0}{\partial \varphi_c} - \frac{\partial N}{\partial \varphi_c} \frac{\partial \varphi_0}{\partial p_c} \bigg)^{-1} \; .
\end{align}

As a side note,  our definition of the moment of inertia~\eqref{eq:defI} differs by a conformal factor $A(\varphi_0)$ from the one used in other recent works~~\cite{Pani:2014jra, Yazadjiev:2016pcb, Silva:2014fca, Doneva:2018ouu, Doneva:2016xmf, Popchev:2018fwu}, where the coordinate rescaling   was not taken into account in the frequency $\Omega$. Since in most cases $A(\varphi_0) \simeq 1$, this omission has  a very small effect on the numerical value of the moment of inertia. However, the contribution of this factor to the angular momentum sensitivity is of order one, as can be seen from the definition~\eqref{eq:angularmomsensitivity}.

\subsection{Newtonian limit} \label{sec:newtonian}

As discussed in  Sec.~\ref{sec:theory}, in the weak-field regime we expect the mass sensitivity to vanish,  $s_M \rightarrow 0$, because in that limit the contribution of the gravitational binding energy to the mass (and thus the dependence on $\tilde G$) disappear. 
At first, one could naively think that this should also be the case for the moment of inertia. However, the radius of a star turns out to depend on the local value of $\tilde G$ even in the Newtonian limit, as we will now show.

Let us focus on the equations describing a Newtonian star in the Jordan frame (we ignore the tildes on variables in this section  to improve readability). The continuity  and hydrostatic equilibrium equations yield a system of equations for the mass $m(r)$ and the pressure $p(r)$:
\begin{align}
    \frac{\mathrm{d} m}{\mathrm{d} r} &= 4 \pi r^2 \rho \label{eq:continuity} \; , \\
    \frac{\mathrm{d}p}{\mathrm{d}r} &= - \frac{G m}{r^2} \rho \label{eq:hydrostatic} \; ,
\end{align}
where $\rho = m_b n_b$ is the energy density of the star, equal to the average baryon mass $m_b$ times the baryon density $n_b$ 
(we are neglecting here the subdominant internal energy density of the fluid).
This system of equations can also be obtained by taking the Newtonian limit of the system~\eqref{eq:EOMmu}-\eqref{eq:EOMkappa} in the Jordan frame.
To close the system,
one needs an EOS, which we choose to be a simple polytrope
\begin{equation} \label{eq:polytropic}
    p = k \rho^{1+1/n} \; ,
\end{equation}
where $n$ is the polytropic index and $k$ is a constant of proportionality. Defining the dimensionless variable $\theta$ by $\rho = \rho_c \theta^n$, where $\rho_c$ is the density at the center of the star, and the dimensionless radius $r = r_0 \xi$ where
\begin{equation}
    r_0^2 = \frac{k(n+1)}{4 \pi G} \rho_c^{1/n-1} \; ,
\end{equation}
one can combine the system~\eqref{eq:continuity}-\eqref{eq:hydrostatic} into a single differential equation on $\theta$, known as the Lane-Emden equation~\cite{1870AmJS...50...57L, 1907gask.book.....E}
\begin{equation}
    \frac{1}{\xi^2} \frac{\mathrm{d}}{\mathrm{d}\xi} \bigg( \xi^2 \frac{\mathrm{d} \theta}{\mathrm{d} \xi} \bigg) + \theta^n =0 \; .
\end{equation}
The radius of this Newtonian star is then defined as the value of $\xi$ for which $\theta$ vanishes, $\theta(\xi_1) = 0$.  To compute the sensitivities, we need to derive the dependence of this radius on the gravitational constant $G$, at fixed total mass. The latter  is given by
\begin{equation}
    M = \int_0^{r_0 \xi_1} 4 \pi r^2 \rho \mathrm{d}r = 4 \pi r_0^3 \rho_c \int_0^{\xi_1} \xi^2 \theta^n \mathrm{d}\xi \; .
\end{equation}
Thus, fixing $M$ yields the relation $\rho_c \propto G^{3n/(3-n)}$. The physical radius of the star is then proportional to $r_0$, and so the moment of inertia has a dependence
\begin{equation}
    \frac{\mathrm{d} \ln I}{\mathrm{d} \ln G} = \frac{2n}{n-3} \; .
\end{equation}
This derivative is  related to the angular momentum sensitivity defined in Eq.~\eqref{eq:angularmomsensitivity} by the dependence of the local value of $G$ on $\varphi_0$ given in Eq.~\eqref{eq:Gtilde}, so that
\begin{equation} \label{eq:sensitivityNewtonian}
    s_I = \frac{1 + \alpha^2 + \alpha'}{1+\alpha^2} \frac{2n}{3-n} \; ,
\end{equation}
where $\alpha' = \mathrm{d}\alpha/\mathrm{d}\varphi$. 
The divergence for $n=3$ can be understood from the fact that a Newtonian polytropic star with an index $n \geq 3$ is unstable~\cite{1983bhwd.book.....S}, and  the moment of inertia changes dramatically as one approaches the instability limit. We will observe this divergent behavior at low neutron star masses in Sec.~\ref{sec:results}, where we present the results of our numerical integration. 

Moreover, Eq.~\eqref{eq:sensitivityNewtonian} suggests that in principle it should  be possible to use Newtonian tests of gravity to constrain scalar-tensor theories, e.g. using measurements involving the moments of inertia, radii or periods of planets. Indeed, this idea has been extensively studied in the past to set bounds on possible time variations of the Newton constant, see e.g.~\cite{RevModPhys.75.403} for a review.

\section{Results} \label{sec:results}

In this section, we will present our numerical results for the mass and angular momentum sensitivities. We integrate the equations of motion~\eqref{eq:EOMmu}-\eqref{eq:EOMkappa} with a fourth-order Runge Kutta integrator, using three tabulated EOS's, \texttt{AP4}, \texttt{SLy} and \texttt{MPA1} (see e.g.~\cite{Read:2008iy} for a discussion of the models). Additionally, we also show the results for a simple polytropic EOS:
\begin{equation}
    \tilde \epsilon = \frac{\tilde p}{\Gamma -1} + \bigg( \frac{\tilde p}{k} \bigg)^{1/\Gamma} \; ,
\end{equation}
with parameters
\begin{equation}
    \Gamma = 2 \; , \quad k = 123 \frac{G^3 M_\odot^2}{c^6} \; .
\end{equation}
This last EOS will allow us to verify the Newtonian limit of the sensitivity $s_I$ by comparing it to the analytic result in Eq.~\eqref{eq:sensitivityNewtonian}.

To ensure the correctness of our numerical results, we have checked that they reproduce  the mass sensitivities reported in Refs.~\cite{PhysRevLett.70.2220, Palenzuela:2013hsa}.
We have also  checked that the relation~\eqref{eq:relqsm} between scalar charge and mass sensitivity is verified by the numerical results. We find that the relation between these two quantities (each computed in an independent way) is indeed satisfied up to numerical relative errors of a few percent.
Finally, we also ensure that our neutron star solutions belong to a stable branch following the prescription described in~\cite{Harada:1998ge}, i.e. we consider solutions where the following stability criterion is satisfied:
\begin{equation} \label{eq:stabilitycondition}
    \frac{\partial N}{\partial p_c} \frac{\partial \varphi_0}{\partial \varphi_c} - \frac{\partial N}{\partial \varphi_c} \frac{\partial \varphi_0}{\partial p_c} >0 \; .
\end{equation}

\subsection{Fierz-Jordan-Brans-Dicke theory}

\begin{figure}
    \centering
    \includegraphics[width=1.0\linewidth]{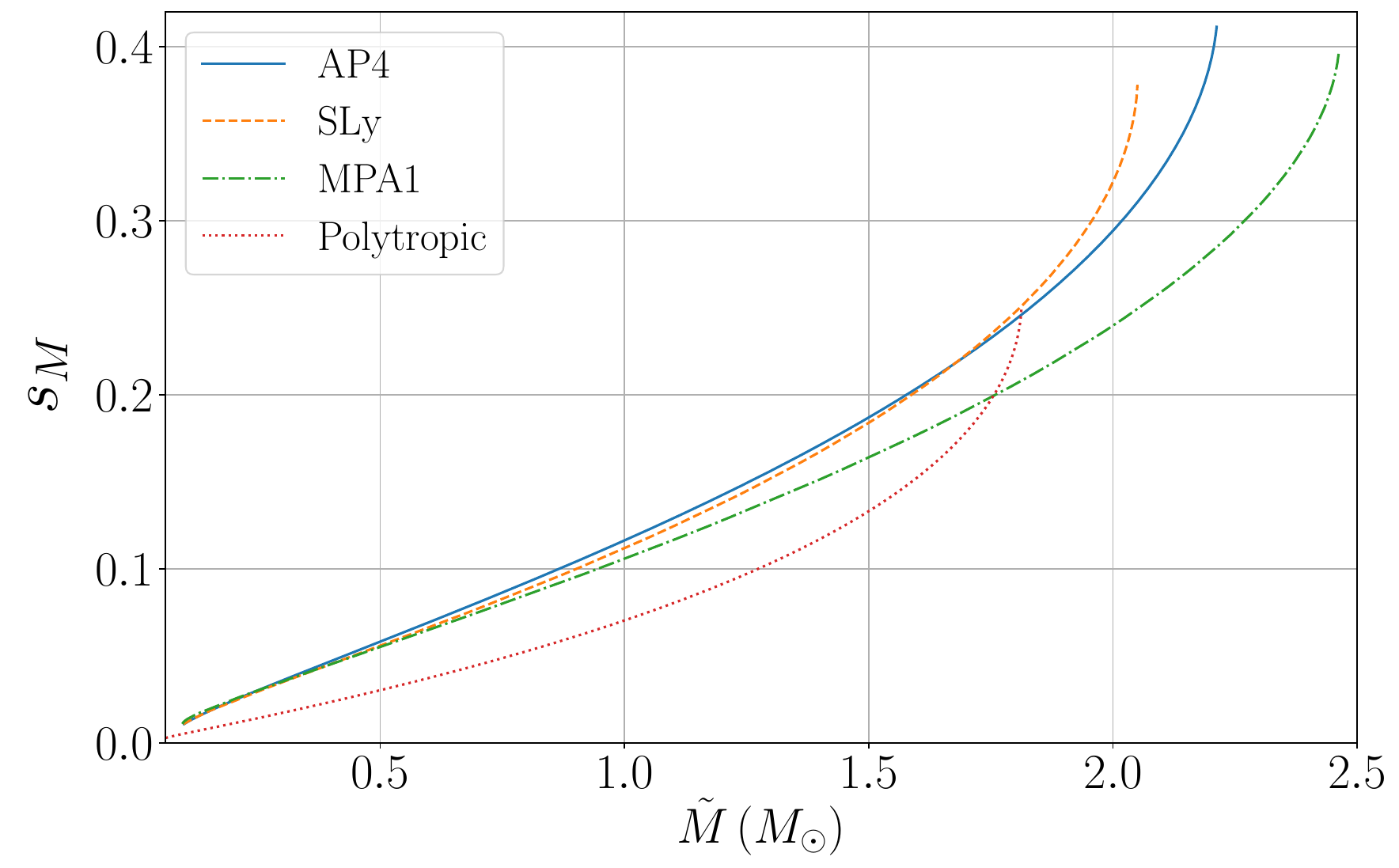}
    \caption{Mass sensitivity $s_M$ as a function of the gravitational mass in the Jordan frame $\tilde M$, for FJBD theory with coupling constant $\alpha = 3 \times 10^{-3}$.}
    \label{fig:sigmaM}
\end{figure}
\begin{figure}
    \centering
    \includegraphics[width=1.0\linewidth]{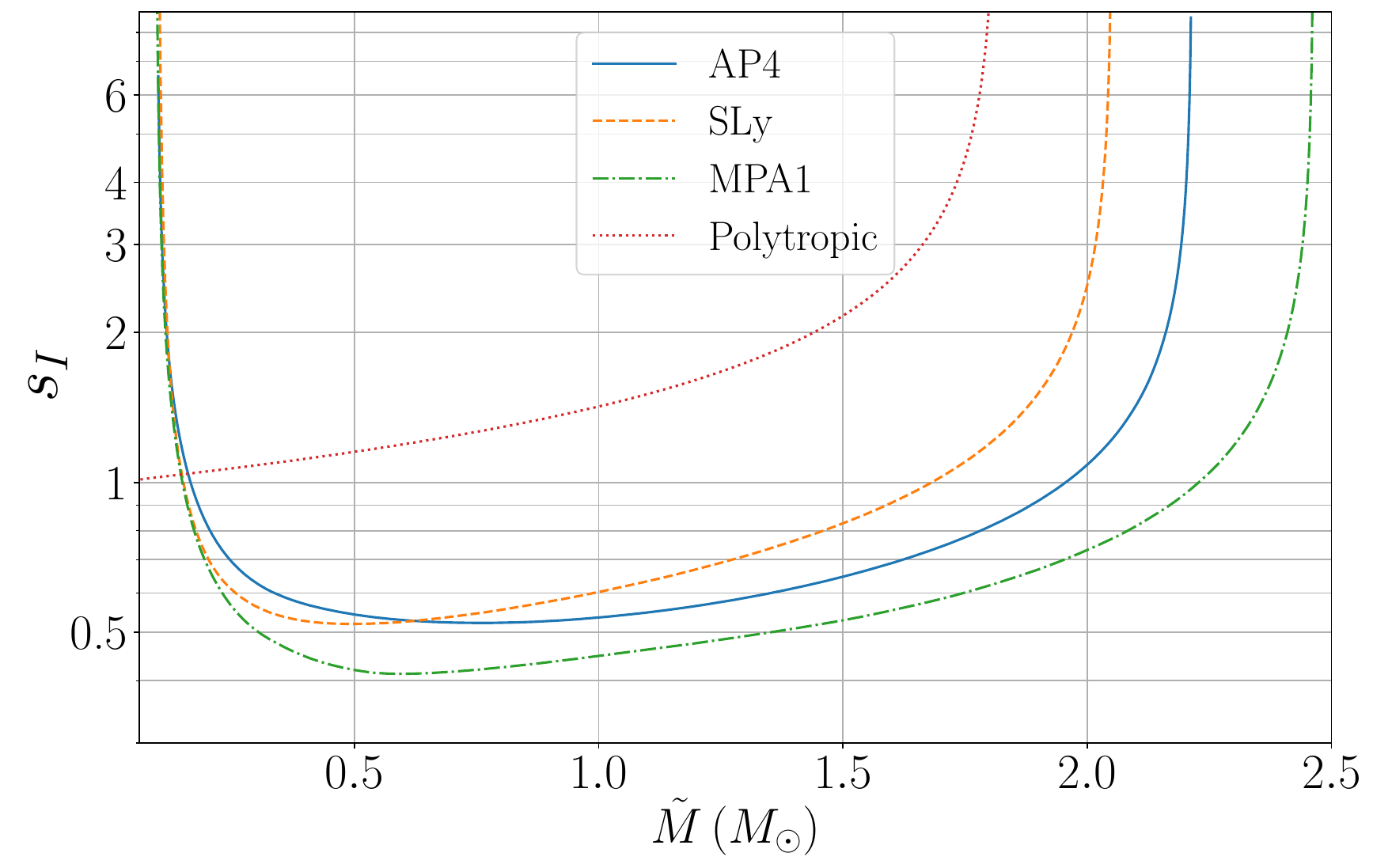}
    \caption{Angular momentum sensitivity $s_I$ as a function of the gravitational mass in the Jordan frame $\tilde M$, for FJBD theory with coupling constant $\alpha = 3 \times 10^{-3}$. The sensitivity diverges at the lower and upper mass ends of the diagram, see main text for more details.}
    \label{fig:sigmaI}
\end{figure}

Our results for the mass sensitivity $s_M$ and the angular momentum sensitivity $s_I$ for FJBD theory are shown in Figs.~\ref{fig:sigmaM} and~\ref{fig:sigmaI}. In this simple theory (and unlike the DEF case), the boundary condition for the scalar field at spatial infinity
bears no physical meaning. Indeed, note that a constant shift of the scalar in the equations of motion~\eqref{eq:EOMmu}-\eqref{eq:EOMkappa} simply translates into a renormalization of the (bare) Newton constant $G_b$. 
Moreover, we find that both the mass and angular momentum sensitivities depend very weakly
on the value of the coupling constant $\alpha$ of the theory. For this reason, we plot results only for a reference value $\alpha = 3 \times 10^{-3}$, which saturates the upper limit allowed by observations (c.f. Sec.~\ref{sec:theory}).

As expected, the mass sensitivity $s_M$ is of order  ${\cal O}(0.1)$ and vanishes in the Newtonian limit (i.e. at low masses). 
However,
from Fig.~\ref{fig:sigmaI}
it is immediately clear that 
for the three tabulated EOSs that we are using, $s_I$ becomes very large (and in fact diverges) both at the upper and lower  ends of the mass
range of stable neutron stars. The maximum mass of a neutron star is of course the well-known Tolman-Oppenheimer-Volkoff  limit~\cite{PhysRev.55.364, PhysRev.55.374}, while the lower mass is associated to inverse $\beta$-decay and other neutronization processes (neutron drip) 
pushing the polytropic index $n$ beyond the maximum value $n=3$ allowed by stability~\cite{1993ApJ...414..717C,2006IJMPA..21.1555B, 2013A&A...560A..48P, 2014NuPhA.921...33B}. 
At these endpoints, the stability condition~\eqref{eq:stabilitycondition} is no longer satisfied, and the slightest perturbation to the star results in large variations of its radius and moment of inertia. 
On the other hand, for the polytropic EOS the Newtonian limit remains finite and exactly corresponds to the limit $s_I = 1$ derived from Eq.~\eqref{eq:sensitivityNewtonian}.

Let us stress that while neutron stars close to the maximum mass can be formed by stellar evolution, population studies reveal that neutron stars form with masses $\tilde M \gtrsim 1 M_\odot$, much above the theoretically allowed minimum mass $\tilde M_\mathrm{min} \sim 0.1-0.2 M_\odot$~\cite{2012ApJ...757...55O, Martinez:2015mya}. Thus, while the divergence of $s_I$ at the lower mass end is an interesting phenomenon, it is not expected to be observationally relevant. Finally, notice that any experimental constraint on the coupling constant $\alpha$ using results from this article will be partly degenerate with changes in the equation of state of nuclear matter, see our companion paper~\cite{Smarra:2024kvv} for more details.

\subsection{Damour and Esposito-Far\`ese theory}

\begin{figure}
    \centering
    \includegraphics[width=1.0\linewidth]{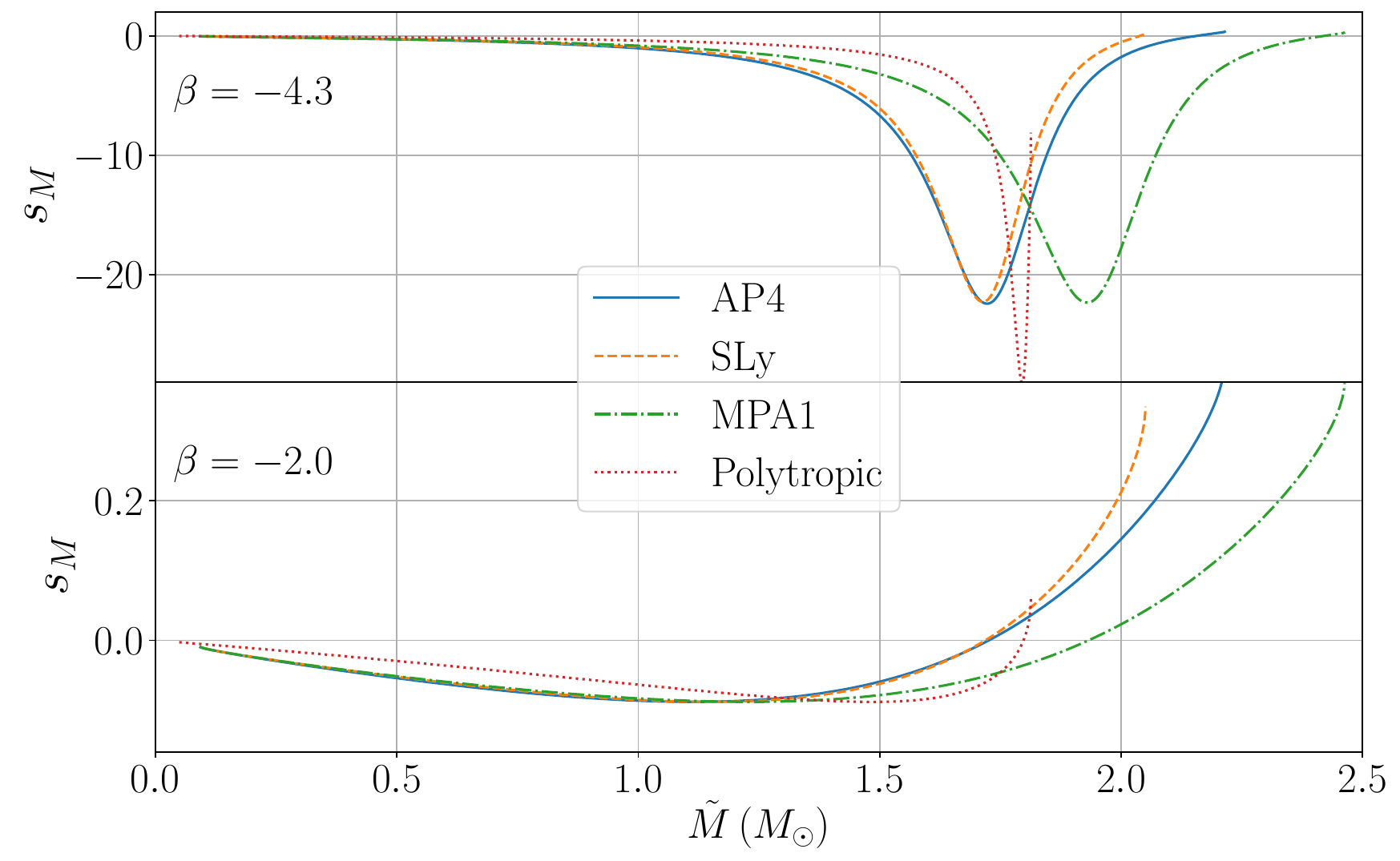}
    \caption{Mass sensitivity $s_M$ as a function of the gravitational mass in the Jordan frame $\tilde M$, for DEF theory. }
    \label{fig:DEF_sigmaM}
\end{figure}

\begin{figure}
    \centering
    \includegraphics[width=1.0\linewidth]{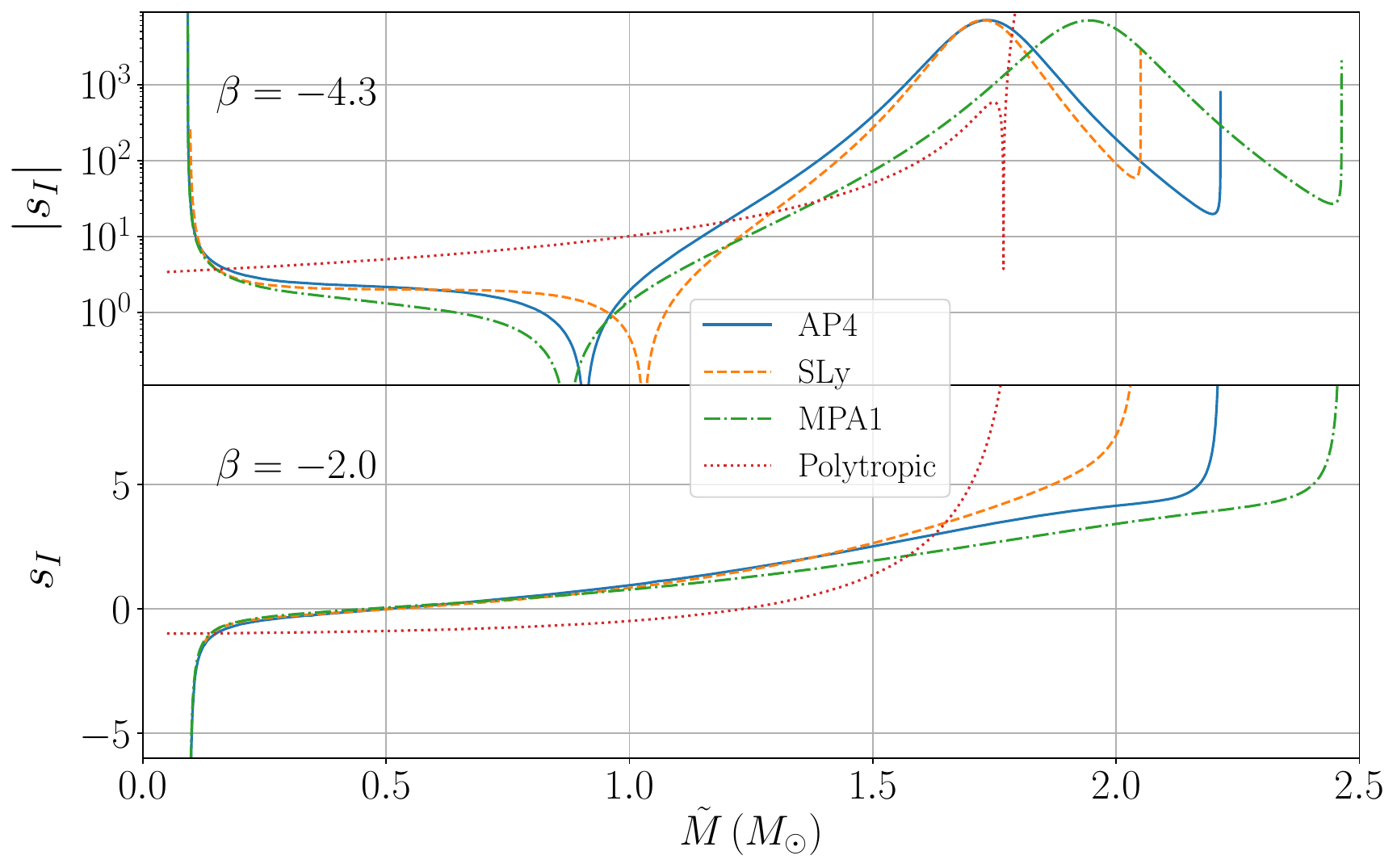}
    \caption{Angular momentum sensitivity $s_I$ as a function of the gravitational mass in the Jordan frame $\tilde M$, for DEF theory. Notice that the absolute value of the sensitivity is plotted in log scale in the upper diagram, however the sensitivity itself is negative in the left part of the plot and positive in its right part.}
    \label{fig:DEF_sigmaI}
\end{figure}

\begin{figure}
    \centering
    \includegraphics[width=1.0\linewidth]{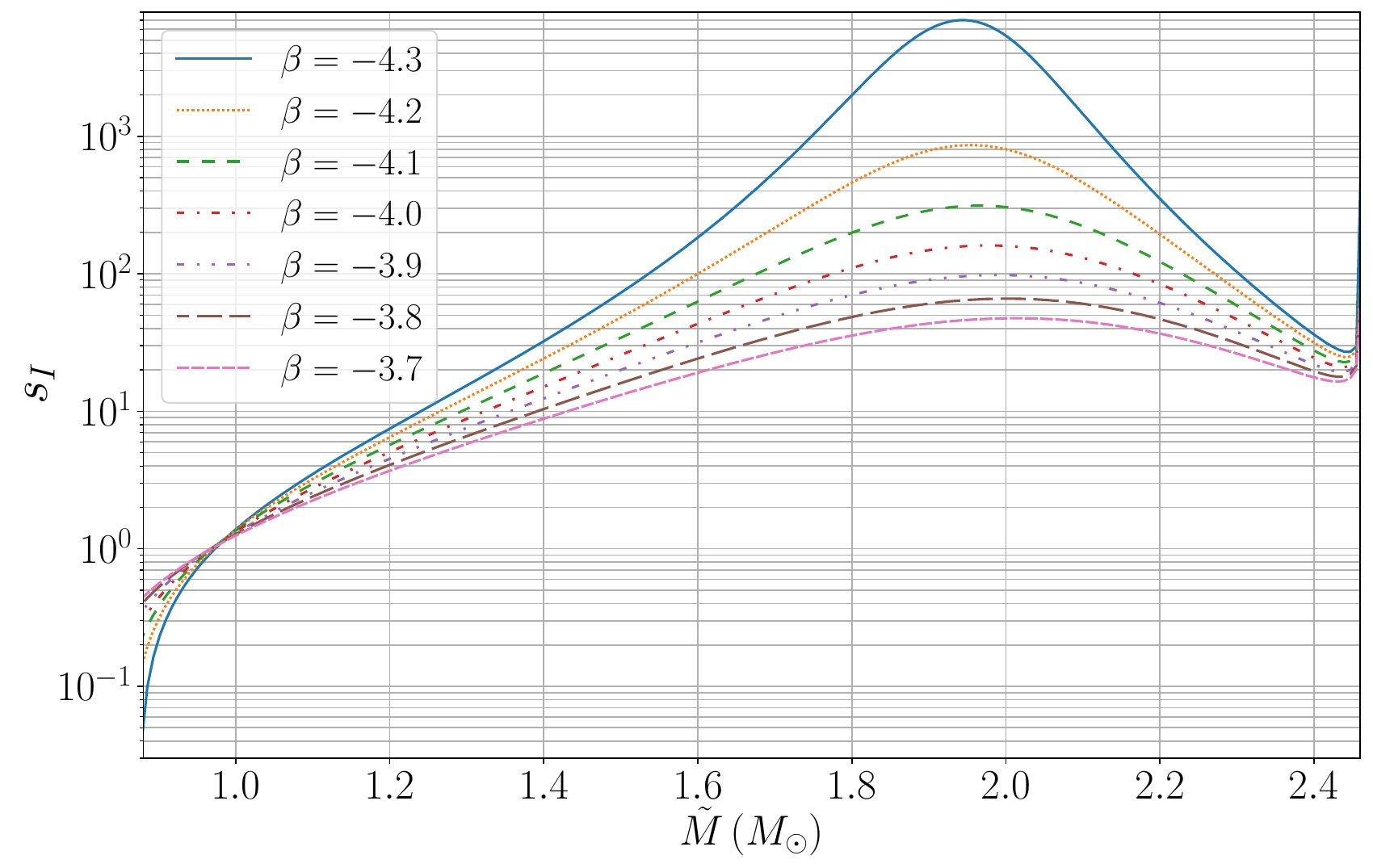}
    \caption{Angular momentum sensitivity $s_I$ as a function of the gravitational mass in the Jordan frame $\tilde M$, for different values of $\beta$ at the onset of spontaneous scalarization.}
    \label{fig:DEF_sigmaI_beta}
\end{figure}

\begin{figure}
    \centering
    \includegraphics[width=1.0\linewidth]{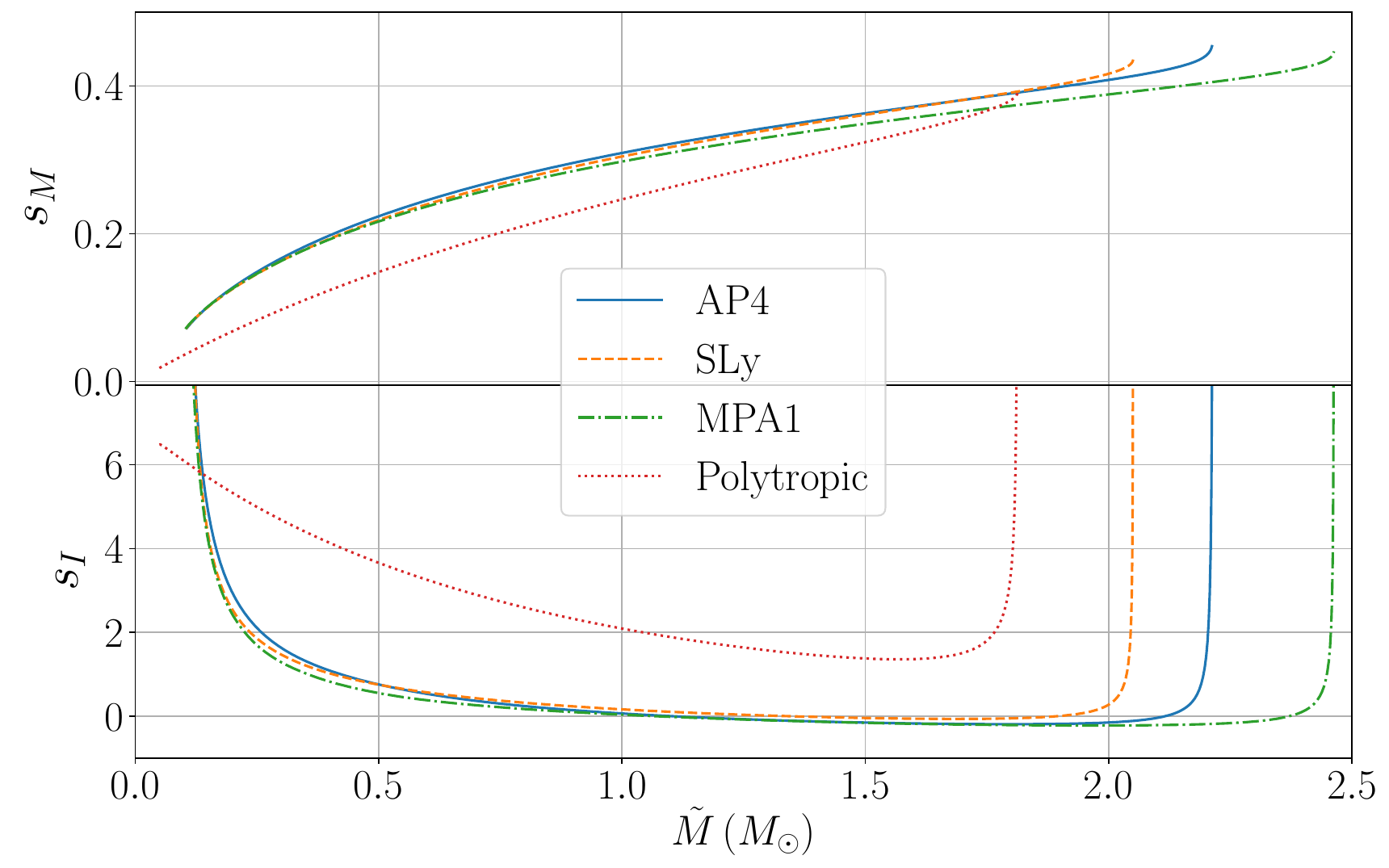}
    \caption{Mass and angular momentum sensitivities as a function of the gravitational mass in the Jordan frame $\tilde M$, for a \textit{positive} value $\beta = 6.0$.}
    \label{fig:DEF_positivebeta}
\end{figure}

Our results for DEF theory are shown in Figs.~\ref{fig:DEF_sigmaM} to~\ref{fig:DEF_positivebeta}. The boundary condition for the scalar at spatial infinity is chosen to be $\varphi_0 = 5 \times 10^{-5}$, which leads to an effective matter coupling $\alpha(\varphi_0) = \beta \varphi_0$ smaller than the experimental bound discussed in Sec.~\ref{sec:theory}. 
We first show in Figs.~\ref{fig:DEF_sigmaM} and~\ref{fig:DEF_sigmaI} the mass and angular momentum sensitivities for two negative values of the parameter $\beta$. The phenomenon of spontaneous scalarization, whereby the sensitivities become large for sufficiently negative values of $\beta \lesssim -4$~\cite{PhysRevLett.70.2220}, is illustrated in the upper panels. The large sensitivity leads to a scalar charge of order one through Eq.~\eqref{eq:relqsm}, even for a vanishingly small asymptotic field $\varphi_0$.
This nonlinear phenomenon explains the strong experimental bounds on these theories in this part of parameter space.
Conversely, for the value $\beta=-2.0$ shown in the lower panel, the sensitivities are of order one or smaller. Therefore, scalar charges are small and  proportional to $\alpha(\varphi_0) = \beta \varphi_0$. In all cases, the sensitivities can change sign, meaning that  masses and moments of inertia can either decrease or increase when varying the strength of gravitational interactions $\tilde G$. This differs from FJBD theory, where both sensitivities did not change sign. 

Quite expectedly, spontaneous scalarization also leads to large angular momentum sensitivities, as can be seen from the upper panel of Fig~\ref{fig:sigmaI}, and from Fig~\ref{fig:DEF_sigmaI_beta}, where we show the dependence of $s_I$ on $\beta$
as spontaneous scalarization gets activated.
Let us give here a quick order-of-magnitude estimate of the constraints that could be obtained on $\beta$ from Pulsar Timing Array (PTA) observations.  When requiring the scalar to be an ultra-light dark matter candidate of mass $m$, its oscillations will produce an additional time residual $\Delta t \sim \beta s_I \varphi(t)^2/m$. However, we know that a similar residual $\Delta t \sim \varphi(t)^2/m$  due to the propagation of the pulsar  signals in a perturbed metric has been constrained in~\cite{Khmelnitsky:2013lxt,EuropeanPulsarTimingArray:2023egv}. By requiring that our additional time residual is smaller than the one constrained in~\cite{Khmelnitsky:2013lxt,EuropeanPulsarTimingArray:2023egv}, we find the constraint $|\beta s_I| \lesssim 1$ in the relevant mass range. A more quantitative analysis is performed in our companion paper~\cite{Smarra:2024kvv}, to which we refer for more details.   

 Additionally,  for the same reason as in FJBD theory, $s_I$ diverges at the upper and lower mass ends for the three realistic EOSs considered. The low-mass (Newtonian) limit for the polytropic EOS is finite and gives a value $s_I \simeq 1+\beta$ in accordance with Eq.~\eqref{eq:sensitivityNewtonian}. 
To further illustrate this point, we plot in Fig.~\ref{fig:DEF_positivebeta} the mass and angular momentum sensitivities for a positive value $\beta = 6.0$.
As can be seen, $s_I$ does not vanish in the Newtonian limit, not even for a 
polytropic EOS (which yields the expected value
 $s_I = 7$ at low masses).

\section{Conclusions}

In this paper, within the context of FJBD and DEF scalar tensor theories, we have developed a formalism to compute the effect of an adiabatic time variation of the scalar field on the moment of inertia and rotational frequency of a slowly rotating neutron star. We have parametrized this effect in terms of angular momentum sensitivities, which we have computed for representative EOSs for nuclear matter, in agreement with the recent constraints from GW170817. 
Our model encompasses time variations of the scalar field induced by cosmological effects or by a putative ``light'' mass term, such as, for instance, the one expected in models of ``fuzzy'' Dark Matter.

The most surprising result that we find is that these angular momentum sensitivities, unlike the mass sensitivities routinely computed and used to constrain scalar tensor theories, do not vanish in the Newtonian limit. We have verified this result analytically using the scaling properties of the Lane-Emden equation. This also allowed us  to understand the divergence of the angular momentum sensitivities for stars that are marginally stable, at the upper and lower end of the neutron star mass range.
For stable neutron stars of masses in the astrophysical range $M\gtrsim 1 M_\odot$, the angular momentum sensitivities are finite. Their values are nearly of ${\cal O}(1)$ in FJBD theory, and get amplified even further in DEF theory, especially for couplings giving rise to spontaneous scalarization.

Among the possible avenues to test our theoretical results, a prominent role is played by millisecond pulsars, for which the rotational frequency is measured very accurately. In a companion paper~\cite{Smarra:2024kvv}, we constrain possible variations of the latter with the data of the European Pulsar Timing Array experiment~\cite{EPTA:2023sfo}. It remains to be seen if the changes in the moment of inertia that we find may be amplified in the dynamical regime of binary pulsars (c.f. the induced and dynamical scalarization of binaries found in Refs.~\cite{Palenzuela:2013hsa,Barausse_2013}, which generalizes the concept of mass sensitivities to binaries).

 \begin{acknowledgments}
  We acknowledge support from the European Union's H2020 ERC Consolidator Grant ``GRavity from Astrophysical to Microscopic Scales'' (Grant No. GRAMS-815673), the PRIN 2022 grant ``GUVIRP - Gravity tests in the UltraViolet and InfraRed with Pulsar timing'', and the EU Horizon 2020 Research and Innovation Programme under the Marie Sklodowska-Curie Grant Agreement No. 101007855.
\end{acknowledgments}

\appendix
\section{Sensitivities and scalar charges} \label{app:scalcharge}

In scalar-tensor theories, it is well-known that the scalar charge $q$, defined in Eq.~\eqref{eq:defq}, is related to the log-derivative of the Einstein-frame gravitational mass $M_E$ by 
\begin{equation}\label{eq:chargedMdphi}
    q = \frac{\mathrm{d} \ln M_E }{ \mathrm{d} \varphi_0} \bigg|_N \; ,
\end{equation}
where the baryonic number is kept fixed. This can be demonstrated as follows. In the Einstein frame, let us look at the neutron star from a large distance, where its effect can be approximated by that of a point particle with action
\begin{equation}
    S_m = - \int \mathrm{d}\tau \, M_E(\varphi) \; ,
\end{equation}
where $\mathrm{d}\tau = \sqrt{-g_{\mu \nu} \mathrm{d}x^\mu \mathrm{d}x^\nu}$ is the proper time, $\varphi$ is the scalar field at the location of the particle (corresponding to the boundary condition $\varphi_0$ of the ``ultraviolet theory, where the structure of the neutron star is resolved), and $M_E$ is the inertial mass in Einstein frame, which turns out to be the same than gravitational mass in this frame. 

Let us now place the particle at rest at the origin. The equation of motion from the scalar field follows from the action and reads
\begin{equation}
    \Box \varphi = 4 \pi G_b M_E'(\varphi) \delta^3(\bm x) \; .
\end{equation}
An important remark is that the derivative of the gravitational mass should be taken 
while keeping the baryonic number constant, since the latter should always be conserved, i.e.
 $M_E'(\varphi) = \mathrm{d} M_E / \mathrm{d} \varphi \big|_N$. Solving the equation of motion at large distances in the static case, one then gets
\begin{equation}\label{falloff}
    \varphi = \varphi_0 - \frac{ G_b M_E'(\varphi_0)}{r} + \mathcal{O} \bigg( \frac{1}{r^2} \bigg) \; ,
\end{equation}
which proves Eq.~\eqref{eq:chargedMdphi}.

One can easily check this  property of the scalar charge using  numerical results, since one can independently compute $q$ from the asymptotic decay of the scalar field and from numerical
 derivatives of the Einstein-frame mass $M_E$. We find indeed that this relation is satisfied up to relative numerical error of a few percent.
Finally, recalling that the inertial mass is given by $M = M_E/A(\varphi_0)$, one can also easily derive Eq.~\eqref{eq:relqsm} from Eq.~\eqref{eq:chargedMdphi}.

\bibliography{refs3}

\begin{thebibliography}{71}%
\makeatletter
\providecommand \@ifxundefined [1]{%
 \@ifx{#1\undefined}
}%
\providecommand \@ifnum [1]{%
 \ifnum #1\expandafter \@firstoftwo
 \else \expandafter \@secondoftwo
 \fi
}%
\providecommand \@ifx [1]{%
 \ifx #1\expandafter \@firstoftwo
 \else \expandafter \@secondoftwo
 \fi
}%
\providecommand \natexlab [1]{#1}%
\providecommand \enquote  [1]{``#1''}%
\providecommand \bibnamefont  [1]{#1}%
\providecommand \bibfnamefont [1]{#1}%
\providecommand \citenamefont [1]{#1}%
\providecommand \href@noop [0]{\@secondoftwo}%
\providecommand \href [0]{\begingroup \@sanitize@url \@href}%
\providecommand \@href[1]{\@@startlink{#1}\@@href}%
\providecommand \@@href[1]{\endgroup#1\@@endlink}%
\providecommand \@sanitize@url [0]{\catcode `\\12\catcode `\$12\catcode `\&12\catcode `\#12\catcode `\^12\catcode `\_12\catcode `\%12\relax}%
\providecommand \@@startlink[1]{}%
\providecommand \@@endlink[0]{}%
\providecommand \url  [0]{\begingroup\@sanitize@url \@url }%
\providecommand \@url [1]{\endgroup\@href {#1}{\urlprefix }}%
\providecommand \urlprefix  [0]{URL }%
\providecommand \Eprint [0]{\href }%
\providecommand \doibase [0]{https://doi.org/}%
\providecommand \selectlanguage [0]{\@gobble}%
\providecommand \bibinfo  [0]{\@secondoftwo}%
\providecommand \bibfield  [0]{\@secondoftwo}%
\providecommand \translation [1]{[#1]}%
\providecommand \BibitemOpen [0]{}%
\providecommand \bibitemStop [0]{}%
\providecommand \bibitemNoStop [0]{.\EOS\space}%
\providecommand \EOS [0]{\spacefactor3000\relax}%
\providecommand \BibitemShut  [1]{\csname bibitem#1\endcsname}%
\let\auto@bib@innerbib\@empty
\bibitem [{\citenamefont {Fierz}\ and\ \citenamefont {Pauli}(1939)}]{Fierz:1939ix}%
  \BibitemOpen
  \bibfield  {author} {\bibinfo {author} {\bibfnamefont {M.}~\bibnamefont {Fierz}}\ and\ \bibinfo {author} {\bibfnamefont {W.}~\bibnamefont {Pauli}},\ }\bibfield  {title} {\bibinfo {title} {{On relativistic wave equations for particles of arbitrary spin in an electromagnetic field}},\ }\href {https://doi.org/10.1098/rspa.1939.0140} {\bibfield  {journal} {\bibinfo  {journal} {Proc. Roy. Soc. Lond.}\ }\textbf {\bibinfo {volume} {A173}},\ \bibinfo {pages} {211} (\bibinfo {year} {1939})}\BibitemShut {NoStop}%
\bibitem [{\citenamefont {Jordan}(1959)}]{Jordan:1959eg}%
  \BibitemOpen
  \bibfield  {author} {\bibinfo {author} {\bibfnamefont {P.}~\bibnamefont {Jordan}},\ }\bibfield  {title} {\bibinfo {title} {{The present state of Dirac's cosmological hypothesis}},\ }\href {https://doi.org/10.1007/BF01375155} {\bibfield  {journal} {\bibinfo  {journal} {Z.\ Phys.}\ }\textbf {\bibinfo {volume} {157}},\ \bibinfo {pages} {112} (\bibinfo {year} {1959})}\BibitemShut {NoStop}%
\bibitem [{\citenamefont {Brans}\ and\ \citenamefont {Dicke}(1961)}]{Brans:1961sx}%
  \BibitemOpen
  \bibfield  {author} {\bibinfo {author} {\bibfnamefont {C.}~\bibnamefont {Brans}}\ and\ \bibinfo {author} {\bibfnamefont {R.~H.}\ \bibnamefont {Dicke}},\ }\bibfield  {title} {\bibinfo {title} {{Mach's principle and a relativistic theory of gravitation}},\ }\href {https://doi.org/10.1103/PhysRev.124.925} {\bibfield  {journal} {\bibinfo  {journal} {Phys. Rev.}\ }\textbf {\bibinfo {volume} {124}},\ \bibinfo {pages} {925} (\bibinfo {year} {1961})}\BibitemShut {NoStop}%
\bibitem [{\citenamefont {Dicke}(1962)}]{Dicke62}%
  \BibitemOpen
  \bibfield  {author} {\bibinfo {author} {\bibfnamefont {R.~H.}\ \bibnamefont {Dicke}},\ }\bibfield  {title} {\bibinfo {title} {Mach's principle and invariance under transformation of units},\ }\href {https://doi.org/10.1103/PhysRev.125.2163} {\bibfield  {journal} {\bibinfo  {journal} {Phys. Rev.}\ }\textbf {\bibinfo {volume} {125}},\ \bibinfo {pages} {2163} (\bibinfo {year} {1962})}\BibitemShut {NoStop}%
\bibitem [{\citenamefont {Nordtvedt}(1968{\natexlab{a}})}]{Nordtvedt68}%
  \BibitemOpen
  \bibfield  {author} {\bibinfo {author} {\bibfnamefont {K.}~\bibnamefont {Nordtvedt}},\ }\bibfield  {title} {\bibinfo {title} {Equivalence principle for massive bodies. ii. theory},\ }\href {https://doi.org/10.1103/PhysRev.169.1017} {\bibfield  {journal} {\bibinfo  {journal} {Phys. Rev.}\ }\textbf {\bibinfo {volume} {169}},\ \bibinfo {pages} {1017} (\bibinfo {year} {1968}{\natexlab{a}})}\BibitemShut {NoStop}%
\bibitem [{\citenamefont {{Eardley}}(1975)}]{Eardley1975ApJ}%
  \BibitemOpen
  \bibfield  {author} {\bibinfo {author} {\bibfnamefont {D.~M.}\ \bibnamefont {{Eardley}}},\ }\bibfield  {title} {\bibinfo {title} {{Observable effects of a scalar gravitational field in a binary pulsar}},\ }\href {https://doi.org/10.1086/181744} {\bibfield  {journal} {\bibinfo  {journal} {Astrophysical Journal}\ }\textbf {\bibinfo {volume} {196}},\ \bibinfo {pages} {L59} (\bibinfo {year} {1975})}\BibitemShut {NoStop}%
\bibitem [{\citenamefont {{Will}}\ and\ \citenamefont {{Zaglauer}}(1989)}]{1989ApJ...346..366W}%
  \BibitemOpen
  \bibfield  {author} {\bibinfo {author} {\bibfnamefont {C.~M.}\ \bibnamefont {{Will}}}\ and\ \bibinfo {author} {\bibfnamefont {H.~W.}\ \bibnamefont {{Zaglauer}}},\ }\bibfield  {title} {\bibinfo {title} {{Gravitational Radiation, Close Binary Systems, and the Brans-Dicke Theory of Gravity}},\ }\href {https://doi.org/10.1086/168016} {\bibfield  {journal} {\bibinfo  {journal} {\apj}\ }\textbf {\bibinfo {volume} {346}},\ \bibinfo {pages} {366} (\bibinfo {year} {1989})}\BibitemShut {NoStop}%
\bibitem [{\citenamefont {{Will}}(1977)}]{1977ApJ...214..826W}%
  \BibitemOpen
  \bibfield  {author} {\bibinfo {author} {\bibfnamefont {C.~M.}\ \bibnamefont {{Will}}},\ }\bibfield  {title} {\bibinfo {title} {{Gravitational radiation from binary systems in alternative metric theories of gravity: dipole radiation and the binary pulsar.}},\ }\href {https://doi.org/10.1086/155313} {\bibfield  {journal} {\bibinfo  {journal} {\apj}\ }\textbf {\bibinfo {volume} {214}},\ \bibinfo {pages} {826} (\bibinfo {year} {1977})}\BibitemShut {NoStop}%
\bibitem [{\citenamefont {Damour}\ and\ \citenamefont {Esposito-Farese}(1992)}]{Damour:1992we}%
  \BibitemOpen
  \bibfield  {author} {\bibinfo {author} {\bibfnamefont {T.}~\bibnamefont {Damour}}\ and\ \bibinfo {author} {\bibfnamefont {G.}~\bibnamefont {Esposito-Farese}},\ }\bibfield  {title} {\bibinfo {title} {{Tensor multiscalar theories of gravitation}},\ }\href {https://doi.org/10.1088/0264-9381/9/9/015} {\bibfield  {journal} {\bibinfo  {journal} {Class. Quant. Grav.}\ }\textbf {\bibinfo {volume} {9}},\ \bibinfo {pages} {2093} (\bibinfo {year} {1992})}\BibitemShut {NoStop}%
\bibitem [{\citenamefont {Will}(1993)}]{Will:1993ns}%
  \BibitemOpen
  \bibfield  {author} {\bibinfo {author} {\bibfnamefont {C.~M.}\ \bibnamefont {Will}},\ }\href@noop {} {\emph {\bibinfo {title} {{Theory and experiment in gravitational physics}}}}\ (\bibinfo {year} {1993})\BibitemShut {NoStop}%
\bibitem [{\citenamefont {Nordtvedt}(1968{\natexlab{b}})}]{Nordtvedt_1968}%
  \BibitemOpen
  \bibfield  {author} {\bibinfo {author} {\bibfnamefont {K.}~\bibnamefont {Nordtvedt}},\ }\bibfield  {title} {\bibinfo {title} {Equivalence principle for massive bodies. i. phenomenology},\ }\href {https://doi.org/10.1103/physrev.169.1014} {\bibfield  {journal} {\bibinfo  {journal} {Physical Review}\ }\textbf {\bibinfo {volume} {169}},\ \bibinfo {pages} {1014} (\bibinfo {year} {1968}{\natexlab{b}})}\BibitemShut {NoStop}%
\bibitem [{\citenamefont {Nordtvedt}(1968{\natexlab{c}})}]{Nordtvedt:1968qs}%
  \BibitemOpen
  \bibfield  {author} {\bibinfo {author} {\bibfnamefont {K.}~\bibnamefont {Nordtvedt}},\ }\bibfield  {title} {\bibinfo {title} {{Equivalence Principle for Massive Bodies. 2. Theory}},\ }\href {https://doi.org/10.1103/PhysRev.169.1017} {\bibfield  {journal} {\bibinfo  {journal} {Phys. Rev.}\ }\textbf {\bibinfo {volume} {169}},\ \bibinfo {pages} {1017} (\bibinfo {year} {1968}{\natexlab{c}})}\BibitemShut {NoStop}%
\bibitem [{\citenamefont {Horndeski}(1974)}]{Horndeski:1974wa}%
  \BibitemOpen
  \bibfield  {author} {\bibinfo {author} {\bibfnamefont {G.~W.}\ \bibnamefont {Horndeski}},\ }\bibfield  {title} {\bibinfo {title} {{Second-order scalar-tensor field equations in a four-dimensional space}},\ }\href {https://doi.org/10.1007/BF01807638} {\bibfield  {journal} {\bibinfo  {journal} {Int. J. Theor. Phys.}\ }\textbf {\bibinfo {volume} {10}},\ \bibinfo {pages} {363} (\bibinfo {year} {1974})}\BibitemShut {NoStop}%
\bibitem [{\citenamefont {Gleyzes}\ \emph {et~al.}(2015)\citenamefont {Gleyzes}, \citenamefont {Langlois}, \citenamefont {Piazza},\ and\ \citenamefont {Vernizzi}}]{Gleyzes:2014dya}%
  \BibitemOpen
  \bibfield  {author} {\bibinfo {author} {\bibfnamefont {J.}~\bibnamefont {Gleyzes}}, \bibinfo {author} {\bibfnamefont {D.}~\bibnamefont {Langlois}}, \bibinfo {author} {\bibfnamefont {F.}~\bibnamefont {Piazza}},\ and\ \bibinfo {author} {\bibfnamefont {F.}~\bibnamefont {Vernizzi}},\ }\bibfield  {title} {\bibinfo {title} {{Healthy theories beyond Horndeski}},\ }\href {https://doi.org/10.1103/PhysRevLett.114.211101} {\bibfield  {journal} {\bibinfo  {journal} {Phys. Rev. Lett.}\ }\textbf {\bibinfo {volume} {114}},\ \bibinfo {pages} {211101} (\bibinfo {year} {2015})},\ \Eprint {https://arxiv.org/abs/1404.6495} {arXiv:1404.6495 [hep-th]} \BibitemShut {NoStop}%
\bibitem [{\citenamefont {Ben~Achour}\ \emph {et~al.}(2016)\citenamefont {Ben~Achour}, \citenamefont {Crisostomi}, \citenamefont {Koyama}, \citenamefont {Langlois}, \citenamefont {Noui},\ and\ \citenamefont {Tasinato}}]{BenAchour:2016fzp}%
  \BibitemOpen
  \bibfield  {author} {\bibinfo {author} {\bibfnamefont {J.}~\bibnamefont {Ben~Achour}}, \bibinfo {author} {\bibfnamefont {M.}~\bibnamefont {Crisostomi}}, \bibinfo {author} {\bibfnamefont {K.}~\bibnamefont {Koyama}}, \bibinfo {author} {\bibfnamefont {D.}~\bibnamefont {Langlois}}, \bibinfo {author} {\bibfnamefont {K.}~\bibnamefont {Noui}},\ and\ \bibinfo {author} {\bibfnamefont {G.}~\bibnamefont {Tasinato}},\ }\bibfield  {title} {\bibinfo {title} {{Degenerate higher order scalar-tensor theories beyond Horndeski up to cubic order}},\ }\href {https://doi.org/10.1007/JHEP12(2016)100} {\bibfield  {journal} {\bibinfo  {journal} {JHEP}\ }\textbf {\bibinfo {volume} {12}},\ \bibinfo {pages} {100}},\ \Eprint {https://arxiv.org/abs/1608.08135} {arXiv:1608.08135 [hep-th]} \BibitemShut {NoStop}%
\bibitem [{\citenamefont {Damour}\ and\ \citenamefont {Taylor}(1992)}]{Damour:1991rd}%
  \BibitemOpen
  \bibfield  {author} {\bibinfo {author} {\bibfnamefont {T.}~\bibnamefont {Damour}}\ and\ \bibinfo {author} {\bibfnamefont {J.~H.}\ \bibnamefont {Taylor}},\ }\bibfield  {title} {\bibinfo {title} {{Strong field tests of relativistic gravity and binary pulsars}},\ }\href {https://doi.org/10.1103/PhysRevD.45.1840} {\bibfield  {journal} {\bibinfo  {journal} {Phys. Rev. D}\ }\textbf {\bibinfo {volume} {45}},\ \bibinfo {pages} {1840} (\bibinfo {year} {1992})}\BibitemShut {NoStop}%
\bibitem [{\citenamefont {Weisberg}\ and\ \citenamefont {Taylor}(2004)}]{weisberg2004relativistic}%
  \BibitemOpen
  \bibfield  {author} {\bibinfo {author} {\bibfnamefont {J.~M.}\ \bibnamefont {Weisberg}}\ and\ \bibinfo {author} {\bibfnamefont {J.~H.}\ \bibnamefont {Taylor}},\ }\href@noop {} {\bibinfo {title} {Relativistic binary pulsar b1913+16: Thirty years of observations and analysis}} (\bibinfo {year} {2004}),\ \Eprint {https://arxiv.org/abs/astro-ph/0407149} {arXiv:astro-ph/0407149 [astro-ph]} \BibitemShut {NoStop}%
\bibitem [{\citenamefont {Kramer}\ \emph {et~al.}(2006)\citenamefont {Kramer}, \citenamefont {Stairs}, \citenamefont {Manchester}, \citenamefont {McLaughlin}, \citenamefont {Lyne} \emph {et~al.}}]{Kramer:2006nb}%
  \BibitemOpen
  \bibfield  {author} {\bibinfo {author} {\bibfnamefont {M.}~\bibnamefont {Kramer}}, \bibinfo {author} {\bibfnamefont {I.~H.}\ \bibnamefont {Stairs}}, \bibinfo {author} {\bibfnamefont {R.}~\bibnamefont {Manchester}}, \bibinfo {author} {\bibfnamefont {M.}~\bibnamefont {McLaughlin}}, \bibinfo {author} {\bibfnamefont {A.}~\bibnamefont {Lyne}}, \emph {et~al.},\ }\bibfield  {title} {\bibinfo {title} {{Tests of general relativity from timing the double pulsar}},\ }\href {https://doi.org/10.1126/science.1132305} {\bibfield  {journal} {\bibinfo  {journal} {Science}\ }\textbf {\bibinfo {volume} {314}},\ \bibinfo {pages} {97} (\bibinfo {year} {2006})},\ \Eprint {https://arxiv.org/abs/astro-ph/0609417} {arXiv:astro-ph/0609417 [astro-ph]} \BibitemShut {NoStop}%
\bibitem [{\citenamefont {Abbott}\ \emph {et~al.}(2019{\natexlab{a}})\citenamefont {Abbott} \emph {et~al.}}]{LIGOScientific:2018dkp}%
  \BibitemOpen
  \bibfield  {author} {\bibinfo {author} {\bibfnamefont {B.~P.}\ \bibnamefont {Abbott}} \emph {et~al.} (\bibinfo {collaboration} {LIGO Scientific, Virgo}),\ }\bibfield  {title} {\bibinfo {title} {{Tests of General Relativity with GW170817}},\ }\href {https://doi.org/10.1103/PhysRevLett.123.011102} {\bibfield  {journal} {\bibinfo  {journal} {Phys. Rev. Lett.}\ }\textbf {\bibinfo {volume} {123}},\ \bibinfo {pages} {011102} (\bibinfo {year} {2019}{\natexlab{a}})},\ \Eprint {https://arxiv.org/abs/1811.00364} {arXiv:1811.00364 [gr-qc]} \BibitemShut {NoStop}%
\bibitem [{\citenamefont {Abbott}\ \emph {et~al.}(2019{\natexlab{b}})\citenamefont {Abbott} \emph {et~al.}}]{LIGOScientific:2019fpa}%
  \BibitemOpen
  \bibfield  {author} {\bibinfo {author} {\bibfnamefont {B.~P.}\ \bibnamefont {Abbott}} \emph {et~al.} (\bibinfo {collaboration} {LIGO Scientific, Virgo}),\ }\bibfield  {title} {\bibinfo {title} {{Tests of General Relativity with the Binary Black Hole Signals from the LIGO-Virgo Catalog GWTC-1}},\ }\href@noop {} {\bibfield  {journal} {\bibinfo  {journal} {a}\ } (\bibinfo {year} {2019}{\natexlab{b}})},\ \Eprint {https://arxiv.org/abs/1903.04467} {arXiv:1903.04467 [gr-qc]} \BibitemShut {NoStop}%
\bibitem [{\citenamefont {Abbott}\ \emph {et~al.}(2021)\citenamefont {Abbott} \emph {et~al.}}]{LIGOScientific:2020tif}%
  \BibitemOpen
  \bibfield  {author} {\bibinfo {author} {\bibfnamefont {R.}~\bibnamefont {Abbott}} \emph {et~al.} (\bibinfo {collaboration} {LIGO Scientific, Virgo}),\ }\bibfield  {title} {\bibinfo {title} {{Tests of general relativity with binary black holes from the second LIGO-Virgo gravitational-wave transient catalog}},\ }\href {https://doi.org/10.1103/PhysRevD.103.122002} {\bibfield  {journal} {\bibinfo  {journal} {Phys. Rev. D}\ }\textbf {\bibinfo {volume} {103}},\ \bibinfo {pages} {122002} (\bibinfo {year} {2021})},\ \Eprint {https://arxiv.org/abs/2010.14529} {arXiv:2010.14529 [gr-qc]} \BibitemShut {NoStop}%
\bibitem [{\citenamefont {Damour}\ and\ \citenamefont {Esposito-Far{\`e}se}(1998)}]{Damour_1998}%
  \BibitemOpen
  \bibfield  {author} {\bibinfo {author} {\bibfnamefont {T.}~\bibnamefont {Damour}}\ and\ \bibinfo {author} {\bibfnamefont {G.}~\bibnamefont {Esposito-Far{\`e}se}},\ }\bibfield  {title} {\bibinfo {title} {Gravitational-wave versus binary-pulsar tests of strong-field gravity},\ }\bibfield  {journal} {\bibinfo  {journal} {Physical Review D}\ }\textbf {\bibinfo {volume} {58}},\ \href {https://doi.org/10.1103/physrevd.58.042001} {10.1103/physrevd.58.042001} (\bibinfo {year} {1998})\BibitemShut {NoStop}%
\bibitem [{\citenamefont {Freire}\ \emph {et~al.}(2012)\citenamefont {Freire}, \citenamefont {Wex}, \citenamefont {Esposito-Farese}, \citenamefont {Verbiest}, \citenamefont {Bailes}, \citenamefont {Jacoby}, \citenamefont {Kramer}, \citenamefont {Stairs}, \citenamefont {Antoniadis},\ and\ \citenamefont {Janssen}}]{Freire:2012mg}%
  \BibitemOpen
  \bibfield  {author} {\bibinfo {author} {\bibfnamefont {P.~C.~C.}\ \bibnamefont {Freire}}, \bibinfo {author} {\bibfnamefont {N.}~\bibnamefont {Wex}}, \bibinfo {author} {\bibfnamefont {G.}~\bibnamefont {Esposito-Farese}}, \bibinfo {author} {\bibfnamefont {J.~P.~W.}\ \bibnamefont {Verbiest}}, \bibinfo {author} {\bibfnamefont {M.}~\bibnamefont {Bailes}}, \bibinfo {author} {\bibfnamefont {B.~A.}\ \bibnamefont {Jacoby}}, \bibinfo {author} {\bibfnamefont {M.}~\bibnamefont {Kramer}}, \bibinfo {author} {\bibfnamefont {I.~H.}\ \bibnamefont {Stairs}}, \bibinfo {author} {\bibfnamefont {J.}~\bibnamefont {Antoniadis}},\ and\ \bibinfo {author} {\bibfnamefont {G.~H.}\ \bibnamefont {Janssen}},\ }\bibfield  {title} {\bibinfo {title} {{The relativistic pulsar-white dwarf binary PSR J1738+0333 II. The most stringent test of scalar-tensor gravity}},\ }\href {https://doi.org/10.1111/j.1365-2966.2012.21253.x} {\bibfield  {journal} {\bibinfo  {journal} {Mon. Not. Roy. Astron. Soc.}\ }\textbf {\bibinfo {volume} {423}},\ \bibinfo
  {pages} {3328} (\bibinfo {year} {2012})},\ \Eprint {https://arxiv.org/abs/1205.1450} {arXiv:1205.1450 [astro-ph.GA]} \BibitemShut {NoStop}%
\bibitem [{\citenamefont {Shao}\ \emph {et~al.}(2017)\citenamefont {Shao}, \citenamefont {Sennett}, \citenamefont {Buonanno}, \citenamefont {Kramer},\ and\ \citenamefont {Wex}}]{Shao:2017gwu}%
  \BibitemOpen
  \bibfield  {author} {\bibinfo {author} {\bibfnamefont {L.}~\bibnamefont {Shao}}, \bibinfo {author} {\bibfnamefont {N.}~\bibnamefont {Sennett}}, \bibinfo {author} {\bibfnamefont {A.}~\bibnamefont {Buonanno}}, \bibinfo {author} {\bibfnamefont {M.}~\bibnamefont {Kramer}},\ and\ \bibinfo {author} {\bibfnamefont {N.}~\bibnamefont {Wex}},\ }\bibfield  {title} {\bibinfo {title} {{Constraining nonperturbative strong-field effects in scalar-tensor gravity by combining pulsar timing and laser-interferometer gravitational-wave detectors}},\ }\href {https://doi.org/10.1103/PhysRevX.7.041025} {\bibfield  {journal} {\bibinfo  {journal} {Phys. Rev. X}\ }\textbf {\bibinfo {volume} {7}},\ \bibinfo {pages} {041025} (\bibinfo {year} {2017})},\ \Eprint {https://arxiv.org/abs/1704.07561} {arXiv:1704.07561 [gr-qc]} \BibitemShut {NoStop}%
\bibitem [{\citenamefont {de~Rham}\ \emph {et~al.}(2013)\citenamefont {de~Rham}, \citenamefont {Tolley},\ and\ \citenamefont {Wesley}}]{deRham:2012fw}%
  \BibitemOpen
  \bibfield  {author} {\bibinfo {author} {\bibfnamefont {C.}~\bibnamefont {de~Rham}}, \bibinfo {author} {\bibfnamefont {A.~J.}\ \bibnamefont {Tolley}},\ and\ \bibinfo {author} {\bibfnamefont {D.~H.}\ \bibnamefont {Wesley}},\ }\bibfield  {title} {\bibinfo {title} {{Vainshtein Mechanism in Binary Pulsars}},\ }\href {https://doi.org/10.1103/PhysRevD.87.044025} {\bibfield  {journal} {\bibinfo  {journal} {Phys. Rev. D}\ }\textbf {\bibinfo {volume} {87}},\ \bibinfo {pages} {044025} (\bibinfo {year} {2013})},\ \Eprint {https://arxiv.org/abs/1208.0580} {arXiv:1208.0580 [gr-qc]} \BibitemShut {NoStop}%
\bibitem [{\citenamefont {Dar}\ \emph {et~al.}(2019)\citenamefont {Dar}, \citenamefont {De~Rham}, \citenamefont {Deskins}, \citenamefont {Giblin},\ and\ \citenamefont {Tolley}}]{Dar:2018dra}%
  \BibitemOpen
  \bibfield  {author} {\bibinfo {author} {\bibfnamefont {F.}~\bibnamefont {Dar}}, \bibinfo {author} {\bibfnamefont {C.}~\bibnamefont {De~Rham}}, \bibinfo {author} {\bibfnamefont {J.~T.}\ \bibnamefont {Deskins}}, \bibinfo {author} {\bibfnamefont {J.~T.}\ \bibnamefont {Giblin}},\ and\ \bibinfo {author} {\bibfnamefont {A.~J.}\ \bibnamefont {Tolley}},\ }\bibfield  {title} {\bibinfo {title} {{Scalar Gravitational Radiation from Binaries: Vainshtein Mechanism in Time-dependent Systems}},\ }\href {https://doi.org/10.1088/1361-6382/aaf5e8} {\bibfield  {journal} {\bibinfo  {journal} {Class. Quant. Grav.}\ }\textbf {\bibinfo {volume} {36}},\ \bibinfo {pages} {025008} (\bibinfo {year} {2019})},\ \Eprint {https://arxiv.org/abs/1808.02165} {arXiv:1808.02165 [hep-th]} \BibitemShut {NoStop}%
\bibitem [{\citenamefont {Kuntz}(2019)}]{Kuntz:2019plo}%
  \BibitemOpen
  \bibfield  {author} {\bibinfo {author} {\bibfnamefont {A.}~\bibnamefont {Kuntz}},\ }\bibfield  {title} {\bibinfo {title} {{Two-body potential of Vainshtein screened theories}},\ }\href {https://doi.org/10.1103/PhysRevD.100.024024} {\bibfield  {journal} {\bibinfo  {journal} {Phys. Rev. D}\ }\textbf {\bibinfo {volume} {100}},\ \bibinfo {pages} {024024} (\bibinfo {year} {2019})},\ \Eprint {https://arxiv.org/abs/1905.07340} {arXiv:1905.07340 [gr-qc]} \BibitemShut {NoStop}%
\bibitem [{\citenamefont {Bezares}\ \emph {et~al.}(2022)\citenamefont {Bezares}, \citenamefont {Aguilera-Miret}, \citenamefont {ter Haar}, \citenamefont {Crisostomi}, \citenamefont {Palenzuela},\ and\ \citenamefont {Barausse}}]{Bezares:2021dma}%
  \BibitemOpen
  \bibfield  {author} {\bibinfo {author} {\bibfnamefont {M.}~\bibnamefont {Bezares}}, \bibinfo {author} {\bibfnamefont {R.}~\bibnamefont {Aguilera-Miret}}, \bibinfo {author} {\bibfnamefont {L.}~\bibnamefont {ter Haar}}, \bibinfo {author} {\bibfnamefont {M.}~\bibnamefont {Crisostomi}}, \bibinfo {author} {\bibfnamefont {C.}~\bibnamefont {Palenzuela}},\ and\ \bibinfo {author} {\bibfnamefont {E.}~\bibnamefont {Barausse}},\ }\bibfield  {title} {\bibinfo {title} {{No Evidence of Kinetic Screening in Simulations of Merging Binary Neutron Stars beyond General Relativity}},\ }\href {https://doi.org/10.1103/PhysRevLett.128.091103} {\bibfield  {journal} {\bibinfo  {journal} {Phys. Rev. Lett.}\ }\textbf {\bibinfo {volume} {128}},\ \bibinfo {pages} {091103} (\bibinfo {year} {2022})},\ \Eprint {https://arxiv.org/abs/2107.05648} {arXiv:2107.05648 [gr-qc]} \BibitemShut {NoStop}%
\bibitem [{\citenamefont {Bo\v{s}kovi\'c}\ and\ \citenamefont {Barausse}(2023)}]{Boskovic:2023dqk}%
  \BibitemOpen
  \bibfield  {author} {\bibinfo {author} {\bibfnamefont {M.}~\bibnamefont {Bo\v{s}kovi\'c}}\ and\ \bibinfo {author} {\bibfnamefont {E.}~\bibnamefont {Barausse}},\ }\bibfield  {title} {\bibinfo {title} {{Two-body problem in theories with kinetic screening}},\ }\href {https://doi.org/10.1103/PhysRevD.108.064033} {\bibfield  {journal} {\bibinfo  {journal} {Phys. Rev. D}\ }\textbf {\bibinfo {volume} {108}},\ \bibinfo {pages} {064033} (\bibinfo {year} {2023})},\ \Eprint {https://arxiv.org/abs/2305.07725} {arXiv:2305.07725 [gr-qc]} \BibitemShut {NoStop}%
\bibitem [{\citenamefont {Cheung}\ \emph {et~al.}(2008)\citenamefont {Cheung}, \citenamefont {Creminelli}, \citenamefont {Fitzpatrick}, \citenamefont {Kaplan},\ and\ \citenamefont {Senatore}}]{Cheung:2007st}%
  \BibitemOpen
  \bibfield  {author} {\bibinfo {author} {\bibfnamefont {C.}~\bibnamefont {Cheung}}, \bibinfo {author} {\bibfnamefont {P.}~\bibnamefont {Creminelli}}, \bibinfo {author} {\bibfnamefont {A.~L.}\ \bibnamefont {Fitzpatrick}}, \bibinfo {author} {\bibfnamefont {J.}~\bibnamefont {Kaplan}},\ and\ \bibinfo {author} {\bibfnamefont {L.}~\bibnamefont {Senatore}},\ }\bibfield  {title} {\bibinfo {title} {{The Effective Field Theory of Inflation}},\ }\href {https://doi.org/10.1088/1126-6708/2008/03/014} {\bibfield  {journal} {\bibinfo  {journal} {JHEP}\ }\textbf {\bibinfo {volume} {03}},\ \bibinfo {pages} {014}},\ \Eprint {https://arxiv.org/abs/0709.0293} {arXiv:0709.0293 [hep-th]} \BibitemShut {NoStop}%
\bibitem [{\citenamefont {Gubitosi}\ \emph {et~al.}(2013)\citenamefont {Gubitosi}, \citenamefont {Piazza},\ and\ \citenamefont {Vernizzi}}]{Gubitosi:2012hu}%
  \BibitemOpen
  \bibfield  {author} {\bibinfo {author} {\bibfnamefont {G.}~\bibnamefont {Gubitosi}}, \bibinfo {author} {\bibfnamefont {F.}~\bibnamefont {Piazza}},\ and\ \bibinfo {author} {\bibfnamefont {F.}~\bibnamefont {Vernizzi}},\ }\bibfield  {title} {\bibinfo {title} {{The Effective Field Theory of Dark Energy}},\ }\href {https://doi.org/10.1088/1475-7516/2013/02/032} {\bibfield  {journal} {\bibinfo  {journal} {JCAP}\ }\textbf {\bibinfo {volume} {1302}},\ \bibinfo {pages} {032}},\ \bibinfo {note} {[JCAP1302,032(2013)]},\ \Eprint {https://arxiv.org/abs/1210.0201} {arXiv:1210.0201 [hep-th]} \BibitemShut {NoStop}%
\bibitem [{\citenamefont {Hui}\ \emph {et~al.}(2017)\citenamefont {Hui}, \citenamefont {Ostriker}, \citenamefont {Tremaine},\ and\ \citenamefont {Witten}}]{Hui:2016ltb}%
  \BibitemOpen
  \bibfield  {author} {\bibinfo {author} {\bibfnamefont {L.}~\bibnamefont {Hui}}, \bibinfo {author} {\bibfnamefont {J.~P.}\ \bibnamefont {Ostriker}}, \bibinfo {author} {\bibfnamefont {S.}~\bibnamefont {Tremaine}},\ and\ \bibinfo {author} {\bibfnamefont {E.}~\bibnamefont {Witten}},\ }\bibfield  {title} {\bibinfo {title} {{Ultralight scalars as cosmological dark matter}},\ }\href {https://doi.org/10.1103/PhysRevD.95.043541} {\bibfield  {journal} {\bibinfo  {journal} {Phys. Rev. D}\ }\textbf {\bibinfo {volume} {95}},\ \bibinfo {pages} {043541} (\bibinfo {year} {2017})},\ \Eprint {https://arxiv.org/abs/1610.08297} {arXiv:1610.08297 [astro-ph.CO]} \BibitemShut {NoStop}%
\bibitem [{\citenamefont {Damour}\ and\ \citenamefont {Esposito-Far\`ese}(1993)}]{PhysRevLett.70.2220}%
  \BibitemOpen
  \bibfield  {author} {\bibinfo {author} {\bibfnamefont {T.}~\bibnamefont {Damour}}\ and\ \bibinfo {author} {\bibfnamefont {G.}~\bibnamefont {Esposito-Far\`ese}},\ }\bibfield  {title} {\bibinfo {title} {Nonperturbative strong-field effects in tensor-scalar theories of gravitation},\ }\href {https://doi.org/10.1103/PhysRevLett.70.2220} {\bibfield  {journal} {\bibinfo  {journal} {Phys. Rev. Lett.}\ }\textbf {\bibinfo {volume} {70}},\ \bibinfo {pages} {2220} (\bibinfo {year} {1993})}\BibitemShut {NoStop}%
\bibitem [{\citenamefont {Blas}\ \emph {et~al.}(2017)\citenamefont {Blas}, \citenamefont {Nacir},\ and\ \citenamefont {Sibiryakov}}]{Blas:2016ddr}%
  \BibitemOpen
  \bibfield  {author} {\bibinfo {author} {\bibfnamefont {D.}~\bibnamefont {Blas}}, \bibinfo {author} {\bibfnamefont {D.~L.}\ \bibnamefont {Nacir}},\ and\ \bibinfo {author} {\bibfnamefont {S.}~\bibnamefont {Sibiryakov}},\ }\bibfield  {title} {\bibinfo {title} {{Ultralight Dark Matter Resonates with Binary Pulsars}},\ }\href {https://doi.org/10.1103/PhysRevLett.118.261102} {\bibfield  {journal} {\bibinfo  {journal} {Phys. Rev. Lett.}\ }\textbf {\bibinfo {volume} {118}},\ \bibinfo {pages} {261102} (\bibinfo {year} {2017})},\ \Eprint {https://arxiv.org/abs/1612.06789} {arXiv:1612.06789 [hep-ph]} \BibitemShut {NoStop}%
\bibitem [{\citenamefont {Blas}\ \emph {et~al.}(2020)\citenamefont {Blas}, \citenamefont {L\'opez~Nacir},\ and\ \citenamefont {Sibiryakov}}]{Blas:2019hxz}%
  \BibitemOpen
  \bibfield  {author} {\bibinfo {author} {\bibfnamefont {D.}~\bibnamefont {Blas}}, \bibinfo {author} {\bibfnamefont {D.}~\bibnamefont {L\'opez~Nacir}},\ and\ \bibinfo {author} {\bibfnamefont {S.}~\bibnamefont {Sibiryakov}},\ }\bibfield  {title} {\bibinfo {title} {{Secular effects of ultralight dark matter on binary pulsars}},\ }\href {https://doi.org/10.1103/PhysRevD.101.063016} {\bibfield  {journal} {\bibinfo  {journal} {Phys. Rev. D}\ }\textbf {\bibinfo {volume} {101}},\ \bibinfo {pages} {063016} (\bibinfo {year} {2020})},\ \Eprint {https://arxiv.org/abs/1910.08544} {arXiv:1910.08544 [gr-qc]} \BibitemShut {NoStop}%
\bibitem [{\citenamefont {Khmelnitsky}\ and\ \citenamefont {Rubakov}(2014)}]{Khmelnitsky:2013lxt}%
  \BibitemOpen
  \bibfield  {author} {\bibinfo {author} {\bibfnamefont {A.}~\bibnamefont {Khmelnitsky}}\ and\ \bibinfo {author} {\bibfnamefont {V.}~\bibnamefont {Rubakov}},\ }\bibfield  {title} {\bibinfo {title} {{Pulsar timing signal from ultralight scalar dark matter}},\ }\href {https://doi.org/10.1088/1475-7516/2014/02/019} {\bibfield  {journal} {\bibinfo  {journal} {JCAP}\ }\textbf {\bibinfo {volume} {02}},\ \bibinfo {pages} {019}},\ \Eprint {https://arxiv.org/abs/1309.5888} {arXiv:1309.5888 [astro-ph.CO]} \BibitemShut {NoStop}%
\bibitem [{\citenamefont {Porayko}\ \emph {et~al.}(2018)\citenamefont {Porayko} \emph {et~al.}}]{Porayko:2018sfa}%
  \BibitemOpen
  \bibfield  {author} {\bibinfo {author} {\bibfnamefont {N.~K.}\ \bibnamefont {Porayko}} \emph {et~al.},\ }\bibfield  {title} {\bibinfo {title} {{Parkes Pulsar Timing Array constraints on ultralight scalar-field dark matter}},\ }\href {https://doi.org/10.1103/PhysRevD.98.102002} {\bibfield  {journal} {\bibinfo  {journal} {Phys. Rev. D}\ }\textbf {\bibinfo {volume} {98}},\ \bibinfo {pages} {102002} (\bibinfo {year} {2018})},\ \Eprint {https://arxiv.org/abs/1810.03227} {arXiv:1810.03227 [astro-ph.CO]} \BibitemShut {NoStop}%
\bibitem [{\citenamefont {Smarra}\ \emph {et~al.}(2023)\citenamefont {Smarra} \emph {et~al.}}]{EuropeanPulsarTimingArray:2023egv}%
  \BibitemOpen
  \bibfield  {author} {\bibinfo {author} {\bibfnamefont {C.}~\bibnamefont {Smarra}} \emph {et~al.} (\bibinfo {collaboration} {European Pulsar Timing Array}),\ }\bibfield  {title} {\bibinfo {title} {{Second Data Release from the European Pulsar Timing Array: Challenging the Ultralight Dark Matter Paradigm}},\ }\href {https://doi.org/10.1103/PhysRevLett.131.171001} {\bibfield  {journal} {\bibinfo  {journal} {Phys. Rev. Lett.}\ }\textbf {\bibinfo {volume} {131}},\ \bibinfo {pages} {171001} (\bibinfo {year} {2023})},\ \Eprint {https://arxiv.org/abs/2306.16228} {arXiv:2306.16228 [astro-ph.HE]} \BibitemShut {NoStop}%
\bibitem [{\citenamefont {Afzal}\ \emph {et~al.}(2023)\citenamefont {Afzal} \emph {et~al.}}]{NANOGrav:2023hvm}%
  \BibitemOpen
  \bibfield  {author} {\bibinfo {author} {\bibfnamefont {A.}~\bibnamefont {Afzal}} \emph {et~al.} (\bibinfo {collaboration} {NANOGrav}),\ }\bibfield  {title} {\bibinfo {title} {{The NANOGrav 15 yr Data Set: Search for Signals from New Physics}},\ }\href {https://doi.org/10.3847/2041-8213/acdc91} {\bibfield  {journal} {\bibinfo  {journal} {Astrophys. J. Lett.}\ }\textbf {\bibinfo {volume} {951}},\ \bibinfo {pages} {L11} (\bibinfo {year} {2023})},\ \Eprint {https://arxiv.org/abs/2306.16219} {arXiv:2306.16219 [astro-ph.HE]} \BibitemShut {NoStop}%
\bibitem [{\citenamefont {Smarra}\ \emph {et~al.}(2024)\citenamefont {Smarra} \emph {et~al.}}]{Smarra:2024kvv}%
  \BibitemOpen
  \bibfield  {author} {\bibinfo {author} {\bibfnamefont {C.}~\bibnamefont {Smarra}} \emph {et~al.},\ }\bibfield  {title} {\bibinfo {title} {{Constraints on conformal ultralight dark matter couplings from the European Pulsar Timing Array}},\ }\href@noop {} {\  (\bibinfo {year} {2024})},\ \Eprint {https://arxiv.org/abs/2405.01633} {arXiv:2405.01633 [astro-ph.HE]} \BibitemShut {NoStop}%
\bibitem [{\citenamefont {{Kramer}}\ and\ \citenamefont {{Champion}}(2013)}]{2013CQGra..30v4009K}%
  \BibitemOpen
  \bibfield  {author} {\bibinfo {author} {\bibfnamefont {M.}~\bibnamefont {{Kramer}}}\ and\ \bibinfo {author} {\bibfnamefont {D.~J.}\ \bibnamefont {{Champion}}},\ }\bibfield  {title} {\bibinfo {title} {{The European Pulsar Timing Array and the Large European Array for Pulsars}},\ }\href {https://doi.org/10.1088/0264-9381/30/22/224009} {\bibfield  {journal} {\bibinfo  {journal} {Classical and Quantum Gravity}\ }\textbf {\bibinfo {volume} {30}},\ \bibinfo {eid} {224009} (\bibinfo {year} {2013})}\BibitemShut {NoStop}%
\bibitem [{\citenamefont {Barausse}\ \emph {et~al.}(2013{\natexlab{a}})\citenamefont {Barausse}, \citenamefont {Palenzuela}, \citenamefont {Ponce},\ and\ \citenamefont {Lehner}}]{Barausse:2012da}%
  \BibitemOpen
  \bibfield  {author} {\bibinfo {author} {\bibfnamefont {E.}~\bibnamefont {Barausse}}, \bibinfo {author} {\bibfnamefont {C.}~\bibnamefont {Palenzuela}}, \bibinfo {author} {\bibfnamefont {M.}~\bibnamefont {Ponce}},\ and\ \bibinfo {author} {\bibfnamefont {L.}~\bibnamefont {Lehner}},\ }\bibfield  {title} {\bibinfo {title} {{Neutron-star mergers in scalar-tensor theories of gravity}},\ }\href {https://doi.org/10.1103/PhysRevD.87.081506} {\bibfield  {journal} {\bibinfo  {journal} {Phys. Rev. D}\ }\textbf {\bibinfo {volume} {87}},\ \bibinfo {pages} {081506} (\bibinfo {year} {2013}{\natexlab{a}})},\ \Eprint {https://arxiv.org/abs/1212.5053} {arXiv:1212.5053 [gr-qc]} \BibitemShut {NoStop}%
\bibitem [{\citenamefont {Palenzuela}\ \emph {et~al.}(2014)\citenamefont {Palenzuela}, \citenamefont {Barausse}, \citenamefont {Ponce},\ and\ \citenamefont {Lehner}}]{Palenzuela:2013hsa}%
  \BibitemOpen
  \bibfield  {author} {\bibinfo {author} {\bibfnamefont {C.}~\bibnamefont {Palenzuela}}, \bibinfo {author} {\bibfnamefont {E.}~\bibnamefont {Barausse}}, \bibinfo {author} {\bibfnamefont {M.}~\bibnamefont {Ponce}},\ and\ \bibinfo {author} {\bibfnamefont {L.}~\bibnamefont {Lehner}},\ }\bibfield  {title} {\bibinfo {title} {{Dynamical scalarization of neutron stars in scalar-tensor gravity theories}},\ }\href {https://doi.org/10.1103/PhysRevD.89.044024} {\bibfield  {journal} {\bibinfo  {journal} {Phys. Rev. D}\ }\textbf {\bibinfo {volume} {89}},\ \bibinfo {pages} {044024} (\bibinfo {year} {2014})},\ \Eprint {https://arxiv.org/abs/1310.4481} {arXiv:1310.4481 [gr-qc]} \BibitemShut {NoStop}%
\bibitem [{\citenamefont {Bertotti}\ \emph {et~al.}(2003)\citenamefont {Bertotti}, \citenamefont {Iess},\ and\ \citenamefont {Tortora}}]{Bertotti:2003rm}%
  \BibitemOpen
  \bibfield  {author} {\bibinfo {author} {\bibfnamefont {B.}~\bibnamefont {Bertotti}}, \bibinfo {author} {\bibfnamefont {L.}~\bibnamefont {Iess}},\ and\ \bibinfo {author} {\bibfnamefont {P.}~\bibnamefont {Tortora}},\ }\bibfield  {title} {\bibinfo {title} {{A test of general relativity using radio links with the Cassini spacecraft}},\ }\href {https://doi.org/10.1038/nature01997} {\bibfield  {journal} {\bibinfo  {journal} {Nature}\ }\textbf {\bibinfo {volume} {425}},\ \bibinfo {pages} {374} (\bibinfo {year} {2003})}\BibitemShut {NoStop}%
\bibitem [{\citenamefont {Pani}\ and\ \citenamefont {Berti}(2014)}]{Pani:2014jra}%
  \BibitemOpen
  \bibfield  {author} {\bibinfo {author} {\bibfnamefont {P.}~\bibnamefont {Pani}}\ and\ \bibinfo {author} {\bibfnamefont {E.}~\bibnamefont {Berti}},\ }\bibfield  {title} {\bibinfo {title} {{Slowly rotating neutron stars in scalar-tensor theories}},\ }\href {https://doi.org/10.1103/PhysRevD.90.024025} {\bibfield  {journal} {\bibinfo  {journal} {Phys. Rev. D}\ }\textbf {\bibinfo {volume} {90}},\ \bibinfo {pages} {024025} (\bibinfo {year} {2014})},\ \Eprint {https://arxiv.org/abs/1405.4547} {arXiv:1405.4547 [gr-qc]} \BibitemShut {NoStop}%
\bibitem [{\citenamefont {Yazadjiev}\ \emph {et~al.}(2016)\citenamefont {Yazadjiev}, \citenamefont {Doneva},\ and\ \citenamefont {Popchev}}]{Yazadjiev:2016pcb}%
  \BibitemOpen
  \bibfield  {author} {\bibinfo {author} {\bibfnamefont {S.~S.}\ \bibnamefont {Yazadjiev}}, \bibinfo {author} {\bibfnamefont {D.~D.}\ \bibnamefont {Doneva}},\ and\ \bibinfo {author} {\bibfnamefont {D.}~\bibnamefont {Popchev}},\ }\bibfield  {title} {\bibinfo {title} {{Slowly rotating neutron stars in scalar-tensor theories with a massive scalar field}},\ }\href {https://doi.org/10.1103/PhysRevD.93.084038} {\bibfield  {journal} {\bibinfo  {journal} {Phys. Rev. D}\ }\textbf {\bibinfo {volume} {93}},\ \bibinfo {pages} {084038} (\bibinfo {year} {2016})},\ \Eprint {https://arxiv.org/abs/1602.04766} {arXiv:1602.04766 [gr-qc]} \BibitemShut {NoStop}%
\bibitem [{\citenamefont {Silva}\ \emph {et~al.}(2015)\citenamefont {Silva}, \citenamefont {Macedo}, \citenamefont {Berti},\ and\ \citenamefont {Crispino}}]{Silva:2014fca}%
  \BibitemOpen
  \bibfield  {author} {\bibinfo {author} {\bibfnamefont {H.~O.}\ \bibnamefont {Silva}}, \bibinfo {author} {\bibfnamefont {C.~F.~B.}\ \bibnamefont {Macedo}}, \bibinfo {author} {\bibfnamefont {E.}~\bibnamefont {Berti}},\ and\ \bibinfo {author} {\bibfnamefont {L.~C.~B.}\ \bibnamefont {Crispino}},\ }\bibfield  {title} {\bibinfo {title} {{Slowly rotating anisotropic neutron stars in general relativity and scalar\textendash{}tensor theory}},\ }\href {https://doi.org/10.1088/0264-9381/32/14/145008} {\bibfield  {journal} {\bibinfo  {journal} {Class. Quant. Grav.}\ }\textbf {\bibinfo {volume} {32}},\ \bibinfo {pages} {145008} (\bibinfo {year} {2015})},\ \Eprint {https://arxiv.org/abs/1411.6286} {arXiv:1411.6286 [gr-qc]} \BibitemShut {NoStop}%
\bibitem [{\citenamefont {Doneva}\ \emph {et~al.}(2018)\citenamefont {Doneva}, \citenamefont {Yazadjiev}, \citenamefont {Stergioulas},\ and\ \citenamefont {Kokkotas}}]{Doneva:2018ouu}%
  \BibitemOpen
  \bibfield  {author} {\bibinfo {author} {\bibfnamefont {D.~D.}\ \bibnamefont {Doneva}}, \bibinfo {author} {\bibfnamefont {S.~S.}\ \bibnamefont {Yazadjiev}}, \bibinfo {author} {\bibfnamefont {N.}~\bibnamefont {Stergioulas}},\ and\ \bibinfo {author} {\bibfnamefont {K.~D.}\ \bibnamefont {Kokkotas}},\ }\bibfield  {title} {\bibinfo {title} {{Differentially rotating neutron stars in scalar-tensor theories of gravity}},\ }\href {https://doi.org/10.1103/PhysRevD.98.104039} {\bibfield  {journal} {\bibinfo  {journal} {Phys. Rev. D}\ }\textbf {\bibinfo {volume} {98}},\ \bibinfo {pages} {104039} (\bibinfo {year} {2018})},\ \Eprint {https://arxiv.org/abs/1807.05449} {arXiv:1807.05449 [gr-qc]} \BibitemShut {NoStop}%
\bibitem [{\citenamefont {Doneva}\ and\ \citenamefont {Yazadjiev}(2016)}]{Doneva:2016xmf}%
  \BibitemOpen
  \bibfield  {author} {\bibinfo {author} {\bibfnamefont {D.~D.}\ \bibnamefont {Doneva}}\ and\ \bibinfo {author} {\bibfnamefont {S.~S.}\ \bibnamefont {Yazadjiev}},\ }\bibfield  {title} {\bibinfo {title} {{Rapidly rotating neutron stars with a massive scalar field\textemdash{}structure and universal relations}},\ }\href {https://doi.org/10.1088/1475-7516/2016/11/019} {\bibfield  {journal} {\bibinfo  {journal} {JCAP}\ }\textbf {\bibinfo {volume} {11}},\ \bibinfo {pages} {019}},\ \Eprint {https://arxiv.org/abs/1607.03299} {arXiv:1607.03299 [gr-qc]} \BibitemShut {NoStop}%
\bibitem [{\citenamefont {Popchev}\ \emph {et~al.}(2019)\citenamefont {Popchev}, \citenamefont {Staykov}, \citenamefont {Doneva},\ and\ \citenamefont {Yazadjiev}}]{Popchev:2018fwu}%
  \BibitemOpen
  \bibfield  {author} {\bibinfo {author} {\bibfnamefont {D.}~\bibnamefont {Popchev}}, \bibinfo {author} {\bibfnamefont {K.~V.}\ \bibnamefont {Staykov}}, \bibinfo {author} {\bibfnamefont {D.~D.}\ \bibnamefont {Doneva}},\ and\ \bibinfo {author} {\bibfnamefont {S.~S.}\ \bibnamefont {Yazadjiev}},\ }\bibfield  {title} {\bibinfo {title} {{Moment of inertia\textendash{}mass universal relations for neutron stars in scalar-tensor theory with self-interacting massive scalar field}},\ }\href {https://doi.org/10.1140/epjc/s10052-019-6691-x} {\bibfield  {journal} {\bibinfo  {journal} {Eur. Phys. J. C}\ }\textbf {\bibinfo {volume} {79}},\ \bibinfo {pages} {178} (\bibinfo {year} {2019})},\ \Eprint {https://arxiv.org/abs/1812.00347} {arXiv:1812.00347 [gr-qc]} \BibitemShut {NoStop}%
\bibitem [{\citenamefont {{Goldman}}(1990)}]{1990MNRAS.244..184G}%
  \BibitemOpen
  \bibfield  {author} {\bibinfo {author} {\bibfnamefont {I.}~\bibnamefont {{Goldman}}},\ }\bibfield  {title} {\bibinfo {title} {{Upper limit on G variability derived from the spin-down of PSR 0655 + 64}},\ }\href@noop {} {\bibfield  {journal} {\bibinfo  {journal} {\mnras}\ }\textbf {\bibinfo {volume} {244}},\ \bibinfo {pages} {184} (\bibinfo {year} {1990})}\BibitemShut {NoStop}%
\bibitem [{\citenamefont {Damour}\ and\ \citenamefont {Esposito-Farese}(1996)}]{Damour:1996ke}%
  \BibitemOpen
  \bibfield  {author} {\bibinfo {author} {\bibfnamefont {T.}~\bibnamefont {Damour}}\ and\ \bibinfo {author} {\bibfnamefont {G.}~\bibnamefont {Esposito-Farese}},\ }\bibfield  {title} {\bibinfo {title} {{Tensor - scalar gravity and binary pulsar experiments}},\ }\href {https://doi.org/10.1103/PhysRevD.54.1474} {\bibfield  {journal} {\bibinfo  {journal} {Phys.Rev.}\ }\textbf {\bibinfo {volume} {D54}},\ \bibinfo {pages} {1474} (\bibinfo {year} {1996})},\ \Eprint {https://arxiv.org/abs/gr-qc/9602056} {arXiv:gr-qc/9602056 [gr-qc]} \BibitemShut {NoStop}%
\bibitem [{\citenamefont {{Rezzolla}}\ and\ \citenamefont {{Zanotti}}(2013)}]{2013rehy.book.....R}%
  \BibitemOpen
  \bibfield  {author} {\bibinfo {author} {\bibfnamefont {L.}~\bibnamefont {{Rezzolla}}}\ and\ \bibinfo {author} {\bibfnamefont {O.}~\bibnamefont {{Zanotti}}},\ }\href@noop {} {\emph {\bibinfo {title} {{Relativistic Hydrodynamics}}}}\ (\bibinfo {year} {2013})\BibitemShut {NoStop}%
\bibitem [{\citenamefont {{Hartle}}(1967)}]{1967ApJ...150.1005H}%
  \BibitemOpen
  \bibfield  {author} {\bibinfo {author} {\bibfnamefont {J.~B.}\ \bibnamefont {{Hartle}}},\ }\bibfield  {title} {\bibinfo {title} {{Slowly Rotating Relativistic Stars. I. Equations of Structure}},\ }\href {https://doi.org/10.1086/149400} {\bibfield  {journal} {\bibinfo  {journal} {\apj}\ }\textbf {\bibinfo {volume} {150}},\ \bibinfo {pages} {1005} (\bibinfo {year} {1967})}\BibitemShut {NoStop}%
\bibitem [{\citenamefont {Collaboration}\ and\ \citenamefont {Collaboration}(2017)}]{the_ligo_scientific_collaboration_gw170817:_2017}%
  \BibitemOpen
  \bibfield  {author} {\bibinfo {author} {\bibfnamefont {T.~L.~S.}\ \bibnamefont {Collaboration}}\ and\ \bibinfo {author} {\bibfnamefont {T.~V.}\ \bibnamefont {Collaboration}},\ }\bibfield  {title} {\bibinfo {title} {{GW}170817: {Observation} of {Gravitational} {Waves} from a {Binary} {Neutron} {Star} {Inspiral}},\ }\bibfield  {journal} {\bibinfo  {journal} {Physical Review Letters}\ }\textbf {\bibinfo {volume} {119}},\ \href {https://doi.org/10.1103/PhysRevLett.119.161101} {10.1103/PhysRevLett.119.161101} (\bibinfo {year} {2017}),\ \bibinfo {note} {arXiv: 1710.05832}\BibitemShut {NoStop}%
\bibitem [{\citenamefont {Read}\ \emph {et~al.}(2009)\citenamefont {Read}, \citenamefont {Lackey}, \citenamefont {Owen},\ and\ \citenamefont {Friedman}}]{Read:2008iy}%
  \BibitemOpen
  \bibfield  {author} {\bibinfo {author} {\bibfnamefont {J.~S.}\ \bibnamefont {Read}}, \bibinfo {author} {\bibfnamefont {B.~D.}\ \bibnamefont {Lackey}}, \bibinfo {author} {\bibfnamefont {B.~J.}\ \bibnamefont {Owen}},\ and\ \bibinfo {author} {\bibfnamefont {J.~L.}\ \bibnamefont {Friedman}},\ }\bibfield  {title} {\bibinfo {title} {{Constraints on a phenomenologically parameterized neutron-star equation of state}},\ }\href {https://doi.org/10.1103/PhysRevD.79.124032} {\bibfield  {journal} {\bibinfo  {journal} {Phys. Rev. D}\ }\textbf {\bibinfo {volume} {79}},\ \bibinfo {pages} {124032} (\bibinfo {year} {2009})},\ \Eprint {https://arxiv.org/abs/0812.2163} {arXiv:0812.2163 [astro-ph]} \BibitemShut {NoStop}%
\bibitem [{\citenamefont {{Lane}}(1870)}]{1870AmJS...50...57L}%
  \BibitemOpen
  \bibfield  {author} {\bibinfo {author} {\bibfnamefont {H.~J.}\ \bibnamefont {{Lane}}},\ }\bibfield  {title} {\bibinfo {title} {{On the theoretical temperature of the Sun, under the hypothesis of a gaseous mass maintaining its volume by its internal heat, and depending on the laws of gases as known to terrestrial experiment}},\ }\href {https://doi.org/10.2475/ajs.s2-50.148.57} {\bibfield  {journal} {\bibinfo  {journal} {American Journal of Science}\ }\textbf {\bibinfo {volume} {50}},\ \bibinfo {pages} {57} (\bibinfo {year} {1870})}\BibitemShut {NoStop}%
\bibitem [{\citenamefont {{Emden}}(1907)}]{1907gask.book.....E}%
  \BibitemOpen
  \bibfield  {author} {\bibinfo {author} {\bibfnamefont {R.}~\bibnamefont {{Emden}}},\ }\href@noop {} {\emph {\bibinfo {title} {{Gaskugeln}}}}\ (\bibinfo {year} {1907})\BibitemShut {NoStop}%
\bibitem [{\citenamefont {{Shapiro}}\ and\ \citenamefont {{Teukolsky}}(1983)}]{1983bhwd.book.....S}%
  \BibitemOpen
  \bibfield  {author} {\bibinfo {author} {\bibfnamefont {S.~L.}\ \bibnamefont {{Shapiro}}}\ and\ \bibinfo {author} {\bibfnamefont {S.~A.}\ \bibnamefont {{Teukolsky}}},\ }\href {https://doi.org/10.1002/9783527617661} {\emph {\bibinfo {title} {{Black holes, white dwarfs and neutron stars. The physics of compact objects}}}}\ (\bibinfo {year} {1983})\BibitemShut {NoStop}%
\bibitem [{\citenamefont {Uzan}(2003)}]{RevModPhys.75.403}%
  \BibitemOpen
  \bibfield  {author} {\bibinfo {author} {\bibfnamefont {J.-P.}\ \bibnamefont {Uzan}},\ }\bibfield  {title} {\bibinfo {title} {The fundamental constants and their variation: observational and theoretical status},\ }\href {https://doi.org/10.1103/RevModPhys.75.403} {\bibfield  {journal} {\bibinfo  {journal} {Rev. Mod. Phys.}\ }\textbf {\bibinfo {volume} {75}},\ \bibinfo {pages} {403} (\bibinfo {year} {2003})}\BibitemShut {NoStop}%
\bibitem [{\citenamefont {Harada}(1998)}]{Harada:1998ge}%
  \BibitemOpen
  \bibfield  {author} {\bibinfo {author} {\bibfnamefont {T.}~\bibnamefont {Harada}},\ }\bibfield  {title} {\bibinfo {title} {{Neutron stars in scalar tensor theories of gravity and catastrophe theory}},\ }\href {https://doi.org/10.1103/PhysRevD.57.4802} {\bibfield  {journal} {\bibinfo  {journal} {Phys. Rev. D}\ }\textbf {\bibinfo {volume} {57}},\ \bibinfo {pages} {4802} (\bibinfo {year} {1998})},\ \Eprint {https://arxiv.org/abs/gr-qc/9801049} {arXiv:gr-qc/9801049} \BibitemShut {NoStop}%
\bibitem [{\citenamefont {Tolman}(1939)}]{PhysRev.55.364}%
  \BibitemOpen
  \bibfield  {author} {\bibinfo {author} {\bibfnamefont {R.~C.}\ \bibnamefont {Tolman}},\ }\bibfield  {title} {\bibinfo {title} {Static solutions of einstein's field equations for spheres of fluid},\ }\href {https://doi.org/10.1103/PhysRev.55.364} {\bibfield  {journal} {\bibinfo  {journal} {Phys. Rev.}\ }\textbf {\bibinfo {volume} {55}},\ \bibinfo {pages} {364} (\bibinfo {year} {1939})}\BibitemShut {NoStop}%
\bibitem [{\citenamefont {Oppenheimer}\ and\ \citenamefont {Volkoff}(1939)}]{PhysRev.55.374}%
  \BibitemOpen
  \bibfield  {author} {\bibinfo {author} {\bibfnamefont {J.~R.}\ \bibnamefont {Oppenheimer}}\ and\ \bibinfo {author} {\bibfnamefont {G.~M.}\ \bibnamefont {Volkoff}},\ }\bibfield  {title} {\bibinfo {title} {On massive neutron cores},\ }\href {https://doi.org/10.1103/PhysRev.55.374} {\bibfield  {journal} {\bibinfo  {journal} {Phys. Rev.}\ }\textbf {\bibinfo {volume} {55}},\ \bibinfo {pages} {374} (\bibinfo {year} {1939})}\BibitemShut {NoStop}%
\bibitem [{\citenamefont {{Colpi}}\ \emph {et~al.}(1993)\citenamefont {{Colpi}}, \citenamefont {{Shapiro}},\ and\ \citenamefont {{Teukolsky}}}]{1993ApJ...414..717C}%
  \BibitemOpen
  \bibfield  {author} {\bibinfo {author} {\bibfnamefont {M.}~\bibnamefont {{Colpi}}}, \bibinfo {author} {\bibfnamefont {S.~L.}\ \bibnamefont {{Shapiro}}},\ and\ \bibinfo {author} {\bibfnamefont {S.~A.}\ \bibnamefont {{Teukolsky}}},\ }\bibfield  {title} {\bibinfo {title} {{A Hydrodynamical Model for the Explosion of a Neutron Star Just below the Minimum Mass}},\ }\href {https://doi.org/10.1086/173118} {\bibfield  {journal} {\bibinfo  {journal} {\apj}\ }\textbf {\bibinfo {volume} {414}},\ \bibinfo {pages} {717} (\bibinfo {year} {1993})}\BibitemShut {NoStop}%
\bibitem [{\citenamefont {{Bordbar}}\ and\ \citenamefont {{Hayati}}(2006)}]{2006IJMPA..21.1555B}%
  \BibitemOpen
  \bibfield  {author} {\bibinfo {author} {\bibfnamefont {G.~H.}\ \bibnamefont {{Bordbar}}}\ and\ \bibinfo {author} {\bibfnamefont {M.}~\bibnamefont {{Hayati}}},\ }\bibfield  {title} {\bibinfo {title} {{Computation of Neutron Star Structure Using Modern Equation of State}},\ }\href {https://doi.org/10.1142/S0217751X06028400} {\bibfield  {journal} {\bibinfo  {journal} {International Journal of Modern Physics A}\ }\textbf {\bibinfo {volume} {21}},\ \bibinfo {pages} {1555} (\bibinfo {year} {2006})},\ \Eprint {https://arxiv.org/abs/0810.3482} {arXiv:0810.3482 [astro-ph]} \BibitemShut {NoStop}%
\bibitem [{\citenamefont {{Potekhin}}\ \emph {et~al.}(2013)\citenamefont {{Potekhin}}, \citenamefont {{Fantina}}, \citenamefont {{Chamel}}, \citenamefont {{Pearson}},\ and\ \citenamefont {{Goriely}}}]{2013A&A...560A..48P}%
  \BibitemOpen
  \bibfield  {author} {\bibinfo {author} {\bibfnamefont {A.~Y.}\ \bibnamefont {{Potekhin}}}, \bibinfo {author} {\bibfnamefont {A.~F.}\ \bibnamefont {{Fantina}}}, \bibinfo {author} {\bibfnamefont {N.}~\bibnamefont {{Chamel}}}, \bibinfo {author} {\bibfnamefont {J.~M.}\ \bibnamefont {{Pearson}}},\ and\ \bibinfo {author} {\bibfnamefont {S.}~\bibnamefont {{Goriely}}},\ }\bibfield  {title} {\bibinfo {title} {{Analytical representations of unified equations of state for neutron-star matter}},\ }\href {https://doi.org/10.1051/0004-6361/201321697} {\bibfield  {journal} {\bibinfo  {journal} {\aap}\ }\textbf {\bibinfo {volume} {560}},\ \bibinfo {eid} {A48} (\bibinfo {year} {2013})},\ \Eprint {https://arxiv.org/abs/1310.0049} {arXiv:1310.0049 [astro-ph.SR]} \BibitemShut {NoStop}%
\bibitem [{\citenamefont {{Belvedere}}\ \emph {et~al.}(2014)\citenamefont {{Belvedere}}, \citenamefont {{Boshkayev}}, \citenamefont {{Rueda}},\ and\ \citenamefont {{Ruffini}}}]{2014NuPhA.921...33B}%
  \BibitemOpen
  \bibfield  {author} {\bibinfo {author} {\bibfnamefont {R.}~\bibnamefont {{Belvedere}}}, \bibinfo {author} {\bibfnamefont {K.}~\bibnamefont {{Boshkayev}}}, \bibinfo {author} {\bibfnamefont {J.~A.}\ \bibnamefont {{Rueda}}},\ and\ \bibinfo {author} {\bibfnamefont {R.}~\bibnamefont {{Ruffini}}},\ }\bibfield  {title} {\bibinfo {title} {{Uniformly rotating neutron stars in the global and local charge neutrality cases}},\ }\href {https://doi.org/10.1016/j.nuclphysa.2013.11.001} {\bibfield  {journal} {\bibinfo  {journal} {\nphysa}\ }\textbf {\bibinfo {volume} {921}},\ \bibinfo {pages} {33} (\bibinfo {year} {2014})},\ \Eprint {https://arxiv.org/abs/1307.2836} {arXiv:1307.2836 [astro-ph.SR]} \BibitemShut {NoStop}%
\bibitem [{\citenamefont {{{\"O}zel}}\ \emph {et~al.}(2012)\citenamefont {{{\"O}zel}}, \citenamefont {{Psaltis}}, \citenamefont {{Narayan}},\ and\ \citenamefont {{Santos Villarreal}}}]{2012ApJ...757...55O}%
  \BibitemOpen
  \bibfield  {author} {\bibinfo {author} {\bibfnamefont {F.}~\bibnamefont {{{\"O}zel}}}, \bibinfo {author} {\bibfnamefont {D.}~\bibnamefont {{Psaltis}}}, \bibinfo {author} {\bibfnamefont {R.}~\bibnamefont {{Narayan}}},\ and\ \bibinfo {author} {\bibfnamefont {A.}~\bibnamefont {{Santos Villarreal}}},\ }\bibfield  {title} {\bibinfo {title} {{On the Mass Distribution and Birth Masses of Neutron Stars}},\ }\href {https://doi.org/10.1088/0004-637X/757/1/55} {\bibfield  {journal} {\bibinfo  {journal} {\apj}\ }\textbf {\bibinfo {volume} {757}},\ \bibinfo {eid} {55} (\bibinfo {year} {2012})},\ \Eprint {https://arxiv.org/abs/1201.1006} {arXiv:1201.1006 [astro-ph.HE]} \BibitemShut {NoStop}%
\bibitem [{\citenamefont {Martinez}\ \emph {et~al.}(2015)\citenamefont {Martinez}, \citenamefont {Stovall}, \citenamefont {Freire}, \citenamefont {Deneva}, \citenamefont {Jenet}, \citenamefont {McLaughlin}, \citenamefont {Bagchi}, \citenamefont {Bates},\ and\ \citenamefont {Ridolfi}}]{Martinez:2015mya}%
  \BibitemOpen
  \bibfield  {author} {\bibinfo {author} {\bibfnamefont {J.~G.}\ \bibnamefont {Martinez}}, \bibinfo {author} {\bibfnamefont {K.}~\bibnamefont {Stovall}}, \bibinfo {author} {\bibfnamefont {P.~C.~C.}\ \bibnamefont {Freire}}, \bibinfo {author} {\bibfnamefont {J.~S.}\ \bibnamefont {Deneva}}, \bibinfo {author} {\bibfnamefont {F.~A.}\ \bibnamefont {Jenet}}, \bibinfo {author} {\bibfnamefont {M.~A.}\ \bibnamefont {McLaughlin}}, \bibinfo {author} {\bibfnamefont {M.}~\bibnamefont {Bagchi}}, \bibinfo {author} {\bibfnamefont {S.~D.}\ \bibnamefont {Bates}},\ and\ \bibinfo {author} {\bibfnamefont {A.}~\bibnamefont {Ridolfi}},\ }\bibfield  {title} {\bibinfo {title} {{Pulsar J0453+1559: A Double Neutron Star System with a Large Mass Asymmetry}},\ }\href {https://doi.org/10.1088/0004-637X/812/2/143} {\bibfield  {journal} {\bibinfo  {journal} {Astrophys. J.}\ }\textbf {\bibinfo {volume} {812}},\ \bibinfo {pages} {143} (\bibinfo {year} {2015})},\ \Eprint {https://arxiv.org/abs/1509.08805} {arXiv:1509.08805 [astro-ph.HE]}
  \BibitemShut {NoStop}%
\bibitem [{\citenamefont {Antoniadis}\ \emph {et~al.}(2023)\citenamefont {Antoniadis} \emph {et~al.}}]{EPTA:2023sfo}%
  \BibitemOpen
  \bibfield  {author} {\bibinfo {author} {\bibfnamefont {J.}~\bibnamefont {Antoniadis}} \emph {et~al.} (\bibinfo {collaboration} {EPTA}),\ }\bibfield  {title} {\bibinfo {title} {{The second data release from the European Pulsar Timing Array - I. The dataset and timing analysis}},\ }\href {https://doi.org/10.1051/0004-6361/202346841} {\bibfield  {journal} {\bibinfo  {journal} {Astron. Astrophys.}\ }\textbf {\bibinfo {volume} {678}},\ \bibinfo {pages} {A48} (\bibinfo {year} {2023})},\ \Eprint {https://arxiv.org/abs/2306.16224} {arXiv:2306.16224 [astro-ph.HE]} \BibitemShut {NoStop}%
\bibitem [{\citenamefont {Barausse}\ \emph {et~al.}(2013{\natexlab{b}})\citenamefont {Barausse}, \citenamefont {Palenzuela}, \citenamefont {Ponce},\ and\ \citenamefont {Lehner}}]{Barausse_2013}%
  \BibitemOpen
  \bibfield  {author} {\bibinfo {author} {\bibfnamefont {E.}~\bibnamefont {Barausse}}, \bibinfo {author} {\bibfnamefont {C.}~\bibnamefont {Palenzuela}}, \bibinfo {author} {\bibfnamefont {M.}~\bibnamefont {Ponce}},\ and\ \bibinfo {author} {\bibfnamefont {L.}~\bibnamefont {Lehner}},\ }\bibfield  {title} {\bibinfo {title} {Neutron-star mergers in scalar-tensor theories of gravity},\ }\bibfield  {journal} {\bibinfo  {journal} {Physical Review D}\ }\textbf {\bibinfo {volume} {87}},\ \href {https://doi.org/10.1103/physrevd.87.081506} {10.1103/physrevd.87.081506} (\bibinfo {year} {2013}{\natexlab{b}})\BibitemShut {NoStop}%
\end{thebibliography}%

\end{document}